\documentclass{aastex631}

\usepackage[version=4]{mhchem}
\usepackage{lipsum}
\usepackage{pifont}%
\usepackage{makecell}
\usepackage{tabularx}
\usepackage{physics}
\let\tablenum\relax 
\usepackage{siunitx}

\usepackage{graphicx}
\usepackage{hyperref}

\begin{document}
\title{Biogenic sulfur gases as biosignatures on temperate sub-Neptune waterworlds}

\author[0000-0002-8163-4608]{Shang-Min Tsai}
\affiliation{Department of Earth and Planetary Sciences, University of California, Riverside, CA, USA}

\author[0000-0001-5271-0635]{Hamish Innes}
\affiliation{Atmospheric, Oceanic and Planetary Physics, Department of Physics, University of Oxford, UK}
\affiliation{Freie Universit{\"a}t Berlin, Institute of Geological Sciences, Malteserstrasse 74-100, 12249 Berlin, Germany}

\author[0000-0002-0413-3308]{Nicholas F. Wogan}
\affiliation{Space Science Division, NASA Ames Research Center, Moffett Field, CA 94035}
\affiliation{Virtual Planetary Laboratory, University of Washington, Seattle, WA 98195}

\author[0000-0002-2949-2163]{Edward W. Schwieterman}  
\affiliation{Department of Earth and Planetary Sciences, University of California, Riverside, CA, USA}
\affiliation{Virtual Planetary Laboratory, University of Washington, Seattle, WA 98195}
\affiliation{Blue Marble Space Institute of Science, Seattle, WA, USA}


\begin{abstract}
Theoretical predictions and observational data indicate a class of sub-Neptune exoplanets may have water-rich interiors covered by hydrogen-dominated atmospheres. Provided suitable climate conditions, such planets could host surface liquid oceans. Motivated by recent JWST observations of K2-18 b, we self-consistently model the photochemistry and potential detectability of biogenic sulfur gases in the atmospheres of temperate sub-Neptune waterworlds for the first time. On Earth today, organic sulfur compounds produced by marine biota are rapidly destroyed by photochemical processes before they can accumulate to significant levels. \cite{Shawn2011} suggest that detectable biogenic sulfur signatures could emerge in Archean-like atmospheres with higher biological production or low UV flux.  In this study, we explore biogenic sulfur across a wide range of biological fluxes and stellar UV environments. Critically, the main photochemical sinks are absent on the nightside of tidally locked planets. To address this, we further perform experiments with a 3D GCM and a 2D photochemical model (VULCAN 2D \citep{Tsai2024}) to simulate the global distribution of biogenic gases to investigate their terminator concentrations as seen via transmission spectroscopy. Our models indicate that biogenic sulfur gases can rise to potentially detectable levels on hydrogen-rich waterworlds, but only for enhanced global biosulfur flux ($\gtrsim$20 times modern Earth's flux).  We find that it is challenging to identify DMS at 3.4 $\mu m$ where it strongly overlaps with \ce{CH4}, whereas it is more plausible to detect DMS and companion byproducts, ethylene (\ce{C2H4}) and ethane (\ce{C2H6}), in the mid-infrared between 9 and 13 $\mu m$.
\end{abstract}


\section{Introduction}\label{sec:intro}
The Kepler mission revealed a large fraction of discovered planets with radii between Earth and Neptune \citep{Howard2012,Thompson2018,Bean2021}. Observed masses and radii indicate a population of sub-Neptunes likely containing water-rich interiors, while evolution models also point to the existence of these waterworlds on the population scale \citep{Zeng2019,Venturini2020,Luque2022,Rogers2023}. In particular, several such planets orbiting M dwarf stars receive comparable instellation\footnote{insolation from their host stars} to Earth, placing them in their planetary systems' habitable zones (e.g., Table 1 in \cite{Madhusudhan2021}). These temperate sub-Neptunes have sparked great interest in atmospheric characterization and the assessment of habitability. Given suitable climate conditions \citep{Pierrehumbert2023,Innes2023}, these water-rich worlds could host surface water oceans below hydrogen-rich atmospheres -- recently referred to as ``Hycean'' ({\bf Hy}drogen o{\bf cean}) worlds \citep{Madhusudhan2021,Nixon2021}. Promisingly, their hydrogen-rich envelopes are more favorable for atmospheric characterization in transmission observations due to large scale heights, compared to the higher molecular weight \ce{N2}- or \ce{CO2}-dominated atmospheres.

Since the water detection in the atmosphere of the temperate sub-Neptune K2-18 b \citep{Benneke2019,Tsiaras2019}, follow-up analyses and modeling efforts have sought to unveil the nature of K2-18 b and similar temperate sub-Neptunes. \cite{Madhusudhan2020} explored the bulk composition under different interior structures. \cite{Yu2021,Tsai2021b} proposed using atmospheric compositional evolution to infer the presence of a shallow surface, whereas \cite{Hu2021} investigated the carbon inventories in a Hycean planet scenario. Recent JWST transit observations revealed a strong detection of carbon dioxide (\ce{CO2}) and methane (\ce{CH4}) on K2-18 b \citep{Madhusudhan2023}. Based on this basis and upper limits of carbon monoxide (CO) and ammonia (\ce{NH3}), \cite{Madhusudhan2023} argued that the atmospheric characteristics of K2-18 b are more consistent with a Hycean world scenario. However, several inconsistencies warrant consideration: (1)  It is challenging to accumulate $\%$ level of \ce{CH4} without biological input or thermochemical recycling from the interior \citep{Yu2021,Tsai2021b}. (2) On Hycean planets, CO can still be efficiently produced to around 0.1 -- 1 $\%$ level through photolysis of \ce{CO2}, while the quenched CO abundance is controlled by the interior temperature on Neptune-like planet with a thick ($\gtrsim$ 100 bar) \ce{H2}--atmosphere \citep{Wogan2024}. (3) The cross-correlation method may be required to identify CO in JWST NIRSpec data \citep{Esparza-Borges2023}, while current analysis based on a single transit observation might not be sufficient to conclusively rule out \ce{NH3} before the upcoming MIRI and revisit observations (JWST Cycle 1, GO-2722 and GO-2372). 

Intriguingly, \cite{Madhusudhan2023} also reported a tentative detection of dimethylsulfide (DMS), which is predominately produced by marine microbes on Earth and regarded as a biosignature gas, especially for anoxic biospheres (see \cite{Shawn2011} and \cite{Schwieterman2018} for a review). On modern Earth, DMS is the main biological source of sulfur, about 1.8--3.5 $\times$10$^9$ molecules cm$^{-2}$s$^{-1}$ \citep{Seinfeld2016,Cala2023}. DMS and methanethiol (\ce{CH3SH}) are released from the degradation of dimethylsulfoniopropionate (DMSP), an organosulfur compound produced by marine phytoplankton. Microbial methylation and detoxification processes can also actively produce DMS from hydrogen sulfide (\ce{H2S}) \citep{Li2023}, allowing hypothetical organisms to make use of the \ce{H2S} reservoir in a \ce{H2}-rich environment on Hycean worlds.  In addition, decayed organic substances in the surface ocean can produce carbon disulfide (\ce{CS2}) and carbonyl sulfide (OCS) through photochemical processes.
These biosulfur gases are not directly associated with energy generation but are products or byproducts of physiological responses, referred to as Type III Biosignatures in \cite{Seager2013}. On present-day Earth, these organic sulfur gases are readily destroyed through photolysis or oxidation by OH radicals \citep{Kettle2001,Shawn2011} and thus cannot accumulate to significant concentrations exceeding the ppm level required for remote detection. 

While great uncertainties persist regarding how biology could operate in an \ce{H2}-rich atmosphere, useful analogies can be drawn from anoxic environments on Earth. As early microbial life emerged, the Archean atmosphere was more reducing than present-day, with abundant \ce{CO2} and \ce{CH4} \citep[e.g.,][]{Catling2020}. As indicated by genomic analysis, organic sulfur cycling could date back to the Proterozoic \citep{Mateos2023}, while sulfur-based metabolic pathways likely emerged in Earth’s earliest biosphere in the Archean and may have been common for biological evolution \citep{Pilcher2003,House2003}. On Archean Earth-like worlds, most biogenic sulfur gases are rapidly destroyed in the atmosphere similar to the present-day, except with the major sink being the reaction with atomic O produced by photolysis \citep{Shawn2011}. Curiously, the potential sulfur-based biosignatures and chemical pathways within a \ce{H2}-dominated atmosphere have not yet been fully explored with self-consistent photochemistry. Moreover, many close-in sub-Neptunes around M stars might be tidally locked. On the permanent nightside, such photochemical sinks of these organic sulfur gases do not exist. It is therefore essential to quantify whether the biogenic gases can build up on the terminators where transit observations probe. 


In this Letter, we explore the global distribution and detectability of methane- and sulfur-based biosignature gases on Hycean planets through photochemical and climate modeling. While there is currently no robust evidence that K2-18 b is a Hycean planet, we adopt its properties as a paradigmatic example for Hycean exoplanets overall.

\begin{figure}[ht!]
\epsscale{.9}
\plotone{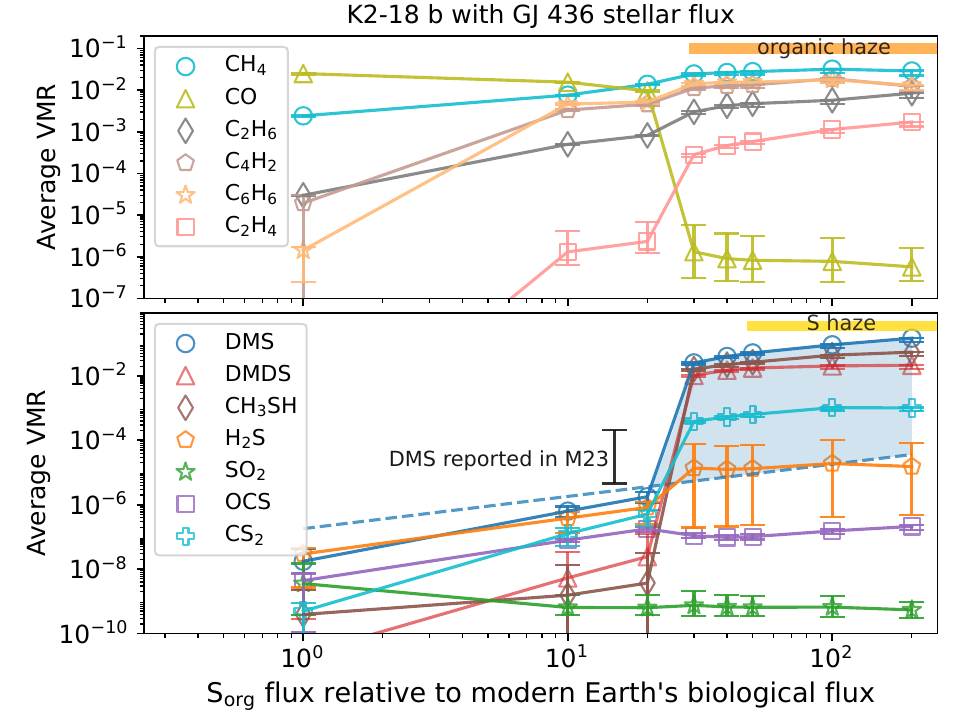}
\plotone{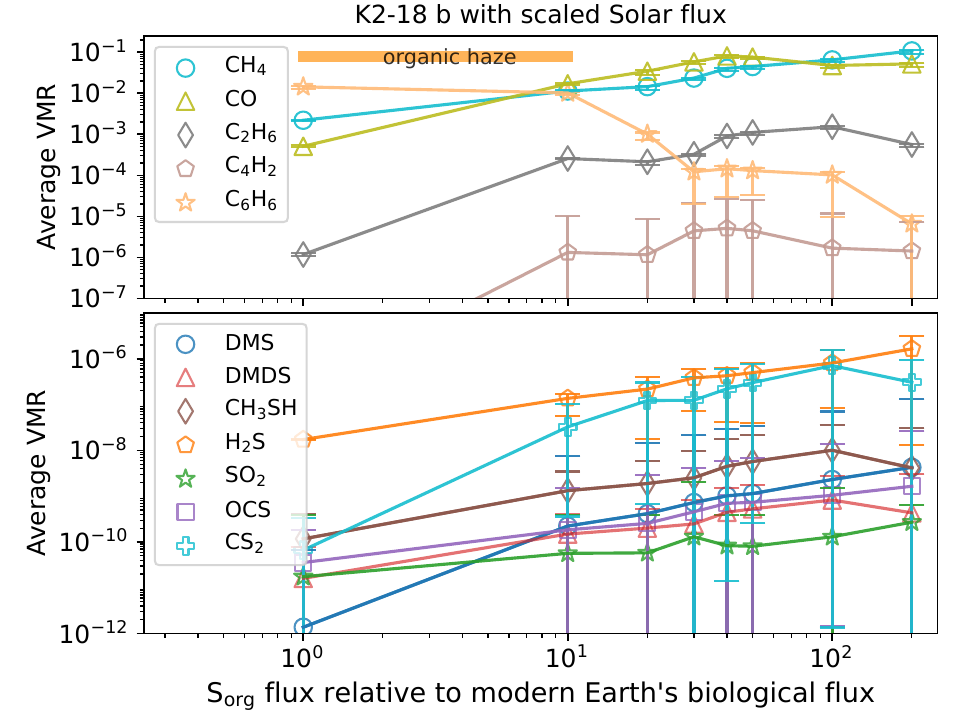}
\caption{The average volume mixing ratios (VMRs) as a function of sulfur biological flux (S$_{\textrm{org}}$). We adopt the stellar spectrum of GJ 436 as an analogous star to K2-18 for our nominal Hycean K2-18 b model (top). Additionally, we scaled the solar flux to match an equivalent flux (bottom). Open circles are the average VMRs across the 1 to 10$^{-4}$ bar pressure range, whereas error bars denote the full span of VMRs within this pressure range. The dashed line represents the limit of DMS given by surface deposition with 0.01 cm$^{-1}$ dry deposition velocity. Combined with the deposition-free DMS abundances in our model, the shaded blue region illustrates the upper and lower bounds of DMS abundance. The black error bar marks the reported DMS VMR from \cite{Madhusudhan2023} (M23). The orange band highlights the regime where hydrocarbon haze precursors, \ce{C4H2} and \ce{C6H6}, exceed $1\%$. Similarly, the yellow band highlights the regime where elemental sulfur (\ce{S8}) becomes saturated and condenses.}\label{fig:DMS-flux}
\end{figure}

\section{Methods}
Here, we model the atmospheric composition of a Hycean planet under a range of biological surface emissions. Using the planetary parameters of K2-18 b as a fiducial example, we vary the surface flux of biosulfur gases and stellar UV flux to quantify the impacts on the global composition of Hycean planet atmospheres.

\subsection{1D climate and photochemical models}
We iterate the radiative transfer (HELIOS; \cite{Malik2019a,Malik2019b}) and photochemical (VULCAN; \cite{Tsai2017,Tsai2021}) models to obtain a self-consistent 1D temperature profile of Hycean K2-18 b. The radiative transfer model HELIOS has been modified to include the moist adiabatic lapse rate in the convective adjustment steps to simulate atmospheres with condensing components\footnote{\url{https://github.com/exoclime/HELIOS/tree/development}}. For our nominal Hycean planet, we assume a 1-bar \ce{H2}--dominated atmosphere with elements given by 100 $\times$ solar metallicity, following \citep{Yu2021,Tsai2021b}. We fix the surface composition of \ce{CO2}, \ce{N2}, and \ce{H2O} in the lower boundary conditions. \ce{CO2} is set to 1$\%$ at the surface, based on the retrieved \ce{CO2} abundance in \cite{Madhusudhan2023} and in accordance with the carbon-inventory estimates in \cite{Kite2018} and \cite{Hu2021}. Surface \ce{N2} is fixed to 0.006, corresponding to all nitrogen (from the 100 $\times$ solar metallicity) partitioning into \ce{N2}. The water vapor is assumed to be saturated above the ocean with a relative humidity of 50$\%$. The spectrum of an analogous M2.5 star GJ 436 is adopted for both HELIOS and VULCAN. 
 
To achieve a marginally habitable surface temperature, we employ an ad hoc surface albedo of 0.3, as previous studies have indicated that highly reflective clouds are necessary to prevent a runaway greenhouse state on K2-18 b \citep{Piette2020,Innes2023}. Similarly, we do not account for the possibility of convection inhibition, which would otherwise raise the surface temperature beyond the supercritical point of water \citep{Innes2023,Pierrehumbert2023}. This allowed us to model a habitable atmospheric state supportive of liquid surface water. The temperature and composition profiles are iterated between HELIOS and VULCAN until the temperature change is about 1$\%$. The planetary parameters for our K2-18 b Hycean model are listed in Table \ref{tab:planet_parameters}. 


The self-consistent temperature and composition structures of our ``lifeless" K2-18 b Hycean planet (excluding biological surface fluxes) are illustrated in Figure \ref{fig:no-S-TP}. We then fed the temperature profile into VULCAN to model the compositional evolution in response to various biological surface fluxes. We updated the S–N–C–H–O photochemical network by revising the reactions involving \ce{H2CO} (see Appendix \ref{app:h2co} and \cite{Wogan2024}) and introduced reactions related to dimethyl sulfide (DMS) and dimethyl disulfide (DMDS)\footnote{\url{https://github.com/exoclime/VULCAN/blob/master/thermo/SNCHO_DMS_photo_network_Tsai2024.txt}}. We consider \ce{CH4} released from methanogenesis and a range of organic sulfur gases biologically emitted from the surface, including \ce{CH3SH}, \ce{CS2}, DMS, dimethyl disulfide (DMDS), and OCS, collectively termed ``S$_{\textrm{org}}$'' after \cite{Shawn2011}. During this stage, the temperature profile from lifeless Hycean is held fixed and we neglect the thermal feedback from the compositional variations due to biological input (mainly \ce{CH4} and S$_{\textrm{org}}$). 

 The rates of gases deposited to the ocean depend on the aerodynamics at the atmosphere-ocean interface and the solubility of the gas \citep{Hu2012,Seinfeld2016}. Due to the uncertainty of parameterizing the dry deposition velocities of several organic sulfur compounds, we assume zero dry deposition for \ce{CH3SH}, DMS, DMDS, and \ce{CS2}, corresponding to the upper limit of biosignature accumulation where biogenic sulfur compounds saturate the oceanic mixed layer. The lower boundary conditions with modern Earth global average S$_{\textrm{org}}$ flux are described in Appendix \ref{app:parameters} and Table \ref{tab:vulcan_parameters}. 



\begin{figure}[ht!]
\includegraphics[width=0.5\textwidth]{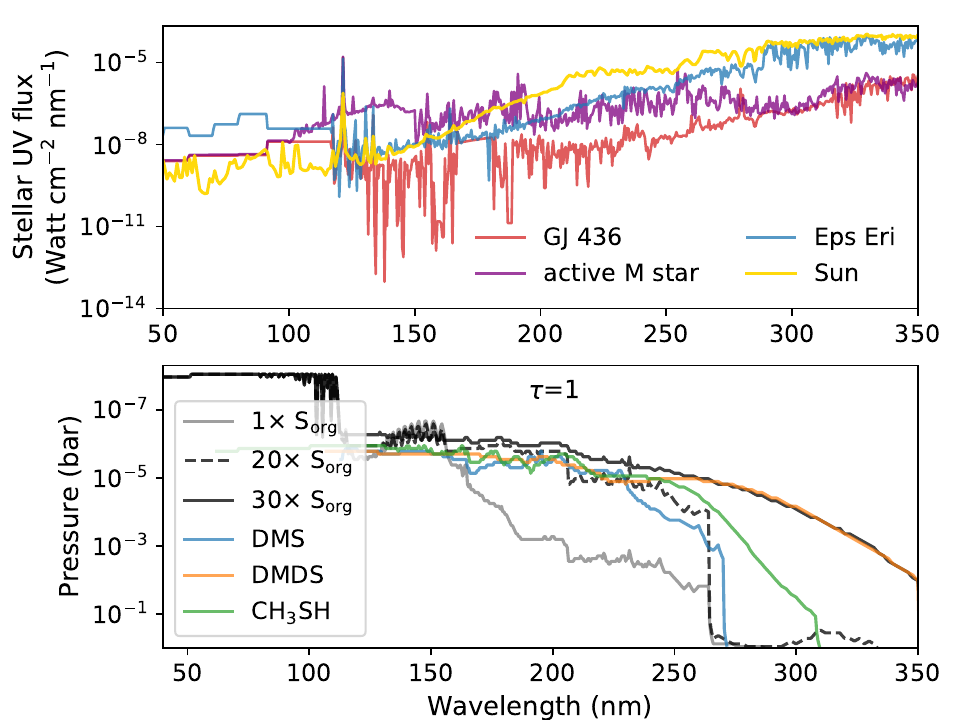}
\includegraphics[width=0.485\textwidth]{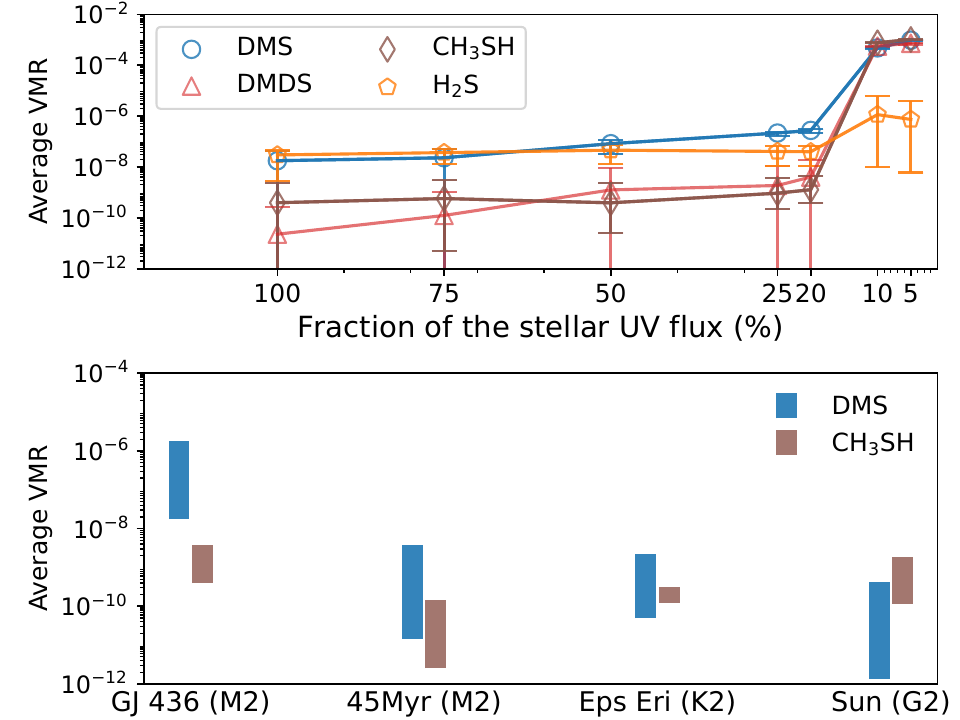}
\caption{Left: The stellar spectra in the UV for GJ 436 (M2.5), a young (45 Myr) and active M star, Epsilon Eridani, and the Sun in the top panel. The spectra of GJ 436 and Epsilon Eridani are from the MUSCLES survey (version 2.2) \citep{France2016,Youngblood2016,Loyd2016}, the active M star spectrum is from HAZMAT \citep{Peacock2020}, and the solar spectrum is from \cite{Gueymard2018}. The lower panel shows the UV photosphere ($\tau$ = 1) with different S$_{\textrm{org}}$. The contributions from DMS, DMDS, and \ce{CH3SH} are further displayed for the 30$\times$ S$_{\textrm{org}}$ model. Right: The upper panel shows the average VMRs of several sulfur species as a function of attenuating total stellar UV flux for Earth's S$_{\textrm{org}}$ flux. The bottom panel shows the span of abundance averaged between  1 and 10$^{-4}$ bar for 1$\times$ -- 20$\times$ S$_{\textrm{org}}$ flux around different stellar types}\label{fig:spectra_tau} 
\end{figure}


\subsection{3D GCM 2D photochemical model}
In the absence of photochemical destruction, biological sulfur compounds can in principle build up on the night side of a tidally-locked Hycean planet. Global circulation plays a central role in transporting species between the dayside and nightside, where distinct chemical sources and sinks are at play \citep{Chen2018,Tsai2021b,Tsai2023c}. To account for the day-night transport, we run VULCAN 2D \citep{Tsai2021b,Tsai2024} to model the equatorial region, using temperature and wind structures from the 3D GCM ExoFMS \citep{Lee2021,Innes2022} as input.

ExoFMS has previously been used to model the climate of a warmer K2-18 b (surface temperature $>$ 500 K) without a liquid ocean and water condensation \citep{Innes2022}. In this study, we fit the shortwave and longwave band opacities in ExoFMS to match our 1D HELIOS correlated-k results. Therefore, the double-gray opacities in our Exo-FMS roughly represent the opacities from the self-consistent 100$\times$ solar metallicity composition. The input parameters are listed in Table \ref{tab:planet_parameters} and the temperature and wind structures can be found in Figure \ref{fig:gcm}. 

The temperatures and winds obtained from exo-FMS are averaged over the equatorial region across 30$^\circ$ and divided into 32 longitude columns to set up VULCAN 2D. In VULCAN 2D, the zonal winds from the GCM are directly implemented for east-west advection, while vertical transport is parameterized by the eddy diffusion coefficient ($K_{\textrm{zz}}$) derived from the mixing length theory with $K_{\textrm{zz}}$ = 0.1 H $\times w_{\textrm{rms}}$, assuming 0.1$\times H$ (scale height) as the characteristic length \citep{Smith1998,Charnay2015}. We use the same lower boundary conditions as the 1D model uniformly across all longitudes.



\section{Results}
\subsection{1D: S$_{\textrm{org}}$ flux $\gtrsim$ 20$\times$ modern Earth value is required to accumulate ppm-levels of DMS}\label{sec:1D}
Figure \ref{fig:DMS-flux} illustrates the average mixing ratios of several carbon-bearing and organic sulfur species in response to increases of biological production of sulfur gases. With modern Earth's S$_{\textrm{org}}$ flux, surface DMS reaches abundances of around 1--30 ppb, before being destroyed above 0.01 bar in our Hycean K2-18 b model. The DMS concentrations on Hycean K2-18 b are slightly higher than modern and Archean Earth's values, which are around 0.01--1 ppb levels \citep{earthDMS,Shawn2011}. As the S$_{\textrm{org}}$ flux continues increasing, there is an abrupt transition between 20--30 times S$_{\textrm{org}}$ flux in our S$_{\textrm{org}}$-deposition-free setup. After this threshold, the atmosphere becomes rich in organic sulfur species and poor in CO, while \ce{CH3SH}, DMS, and DMDS jump from sub-ppm level to above 0.1 $\%$. In the more conservative scenario where the ocean is not saturated, DMS is expected to be limited by the surface deposition and increased linearly with increasing S$_{\textrm{org}}$, as shown by the dashed line in Figure \ref{fig:DMS-flux}. Once S$_{\textrm{org}}$ exceeds 20--30$\times$ Earth's value, \ce{H2S} begins accumulating in the upper atmosphere, as a result of photochemistry reducing sulfur back to its thermochemically favored state. This occurs when \ce{CH3SH} builds up to sufficiently high abundances that its photodissociation becomes significant, producing \ce{H2S} via the following pathway 
\begin{eqnarray}
\begin{aligned}
\ce{CH3SH &->[h\nu] CH3 + SH}\\
\ce{CH3 + SH &-> CH4 + S}\\
\ce{S + H2  &->[M] + H2S}\\
\noalign{\vglue 3pt}
\end{aligned}
\label{re:CH3SH-H2S}
\end{eqnarray}
In fact, the ultimate fate of S$_{\textrm{org}}$ species in a Hycean atmosphere is either photochemical oxidation into \ce{CH3SO} or reduction back to hydrocarbons and \ce{H2S}, where \ce{CH3SO} is mainly produced by DMDS reacting with atomic O. The path to \ce{H2S} becomes more dominant over \ce{CH3SO} as S$_{\textrm{org}}$ flux increases and makes the atmosphere more reducing. Although \ce{SO2} is also photochemically produced, it is highly soluble and deposition into the ocean limits abundances to insignificant levels. 

Notably, the atmosphere also becomes rich in hydrocarbons as S$_{\textrm{org}}$ increases, as pointed out in \cite{Shawn2011}. CO follows an opposite trend to S$_{\textrm{org}}$, showing an abrupt decline after 20 times modern Earth's S$_{\textrm{org}}$ flux. In this carbon-rich regime, the high abundances of haze precursors, such as \ce{C4H2} and \ce{C6H6} \citep{Tsai2021}, imply that the atmosphere would be covered by photochemical hazes \citep{Arney2018}. A main pathway to produce organic haze precursors in the upper atmosphere initiated by DMS photolysis is
\begin{eqnarray}
\begin{aligned}
\ce{DMS (CH3SCH3) &->[h\nu] CH3 + CH3S}\\
\ce{CH3 + CH3 &->[M] C2H6}\\
\ce{C2H6 &->[M] C2H4 + 2H}\\
\ce{C2H4 &->[h\nu] C2H2 + 2H}\\
\ce{C2H2 &->[h\nu] C2H + H}\\
\ce{C2H2 + C2H &-> C4H2 + H}\\
\ce{C4H2 + H &->[M] C4H3}\\
\ce{C2H2 + C4H3 &->[M] C6H5}\\
\ce{C6H5 + H2 &-> C6H6 + H}\\
\noalign{\vglue 3pt}
\end{aligned}
\label{re:DMS-C6H6}
\end{eqnarray}

For DMS at low concentrations below ppm-level, its primary sink is direct photodissociation. After the transition to where DMS has a high concentration, its photodissociation halts in the lower atmosphere due to shielding by DMDS, \ce{CH3SH}, and self-shielding. The reaction with atomic H becomes the main sink. This shielding effect from S$_{\textrm{org}}$ also reduces the photolysis of \ce{CO2} causing the drop of CO when the flux $\gtrsim$ 20$\times$ S$_{\textrm{org}}$. The overall impact of shielding across this transition is evident in the top left panel of Figure \ref{fig:spectra_tau}, where DMS, DMDS, and \ce{CH3SH} significantly elevate the near-UV (NUV) photosphere and subsequently attenuate the NUV flux reaching the surface.  


For an equivalent solar flux, the same average abundances of hydrocarbon and sulfur gases as a function of biological sulfur flux are shown in the bottom of Figure \ref{fig:DMS-flux}. One major difference is that the stronger NUV irradiation generates atomic H down to the surface, which significantly suppresses DMS and other organic sulfur gases. DMS remains below 10 ppb level for S$_{\textrm{org}}$ flux up to 200 times modern Earth's value. Moreover, the concentration of DMS is confined near the surface, whereas for an M-dwarf host star, DMS can extend vertically above 0.01 bar. Since S$_{\textrm{org}}$ never reaches significant abundances under a Sun-like star, shielding of \ce{CO2} photolysis does not occur. Consequently, both \ce{CH4} and CO progressively increase with S$_{\textrm{org}}$ flux. Our model results with an active M star show qualitatively similar trends to those with the equivalent solar flux owing to its enhanced UV flux.


\begin{figure}[ht!]
\includegraphics[width=0.355\textwidth]{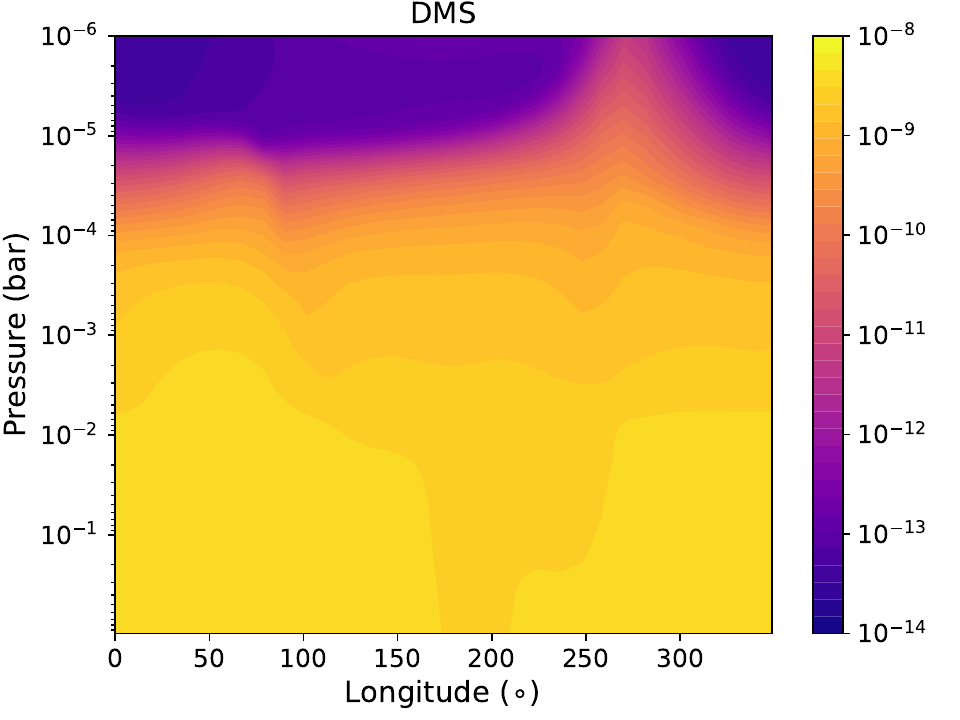}
\includegraphics[width=0.355\textwidth]{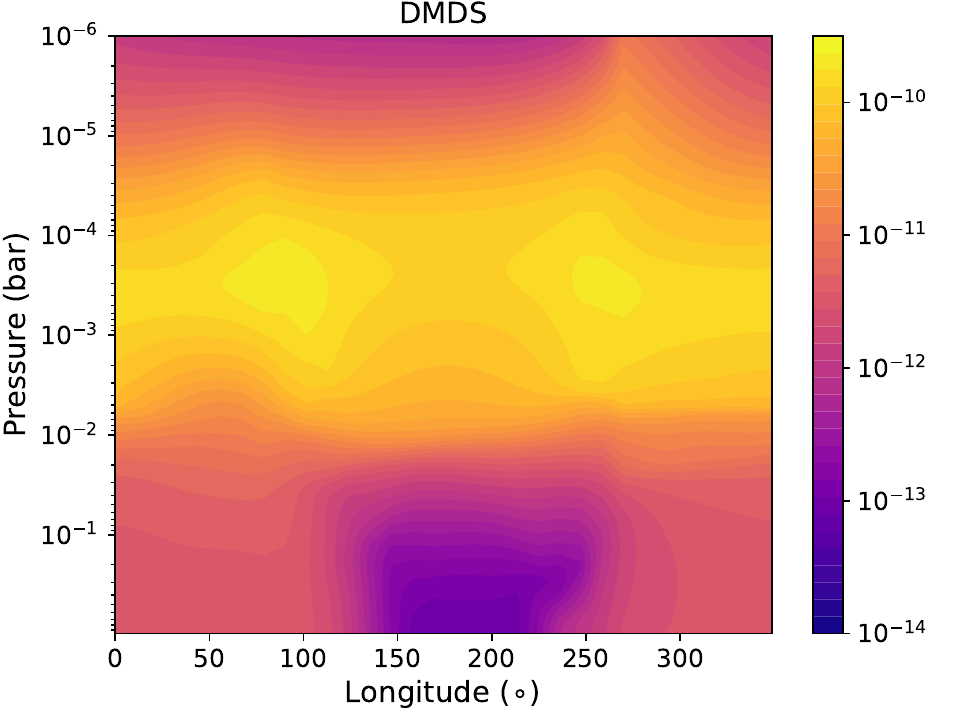}
\includegraphics[width=0.34\textwidth]{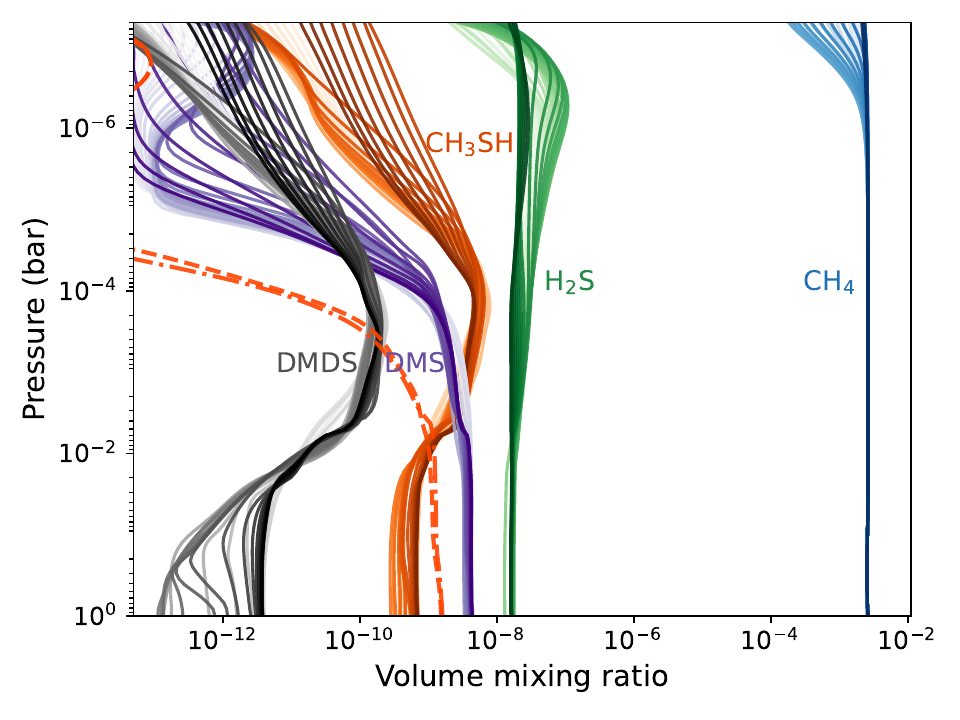}
\caption{The equatorial volume-mixing-ratio distribution of DMS (left) and DMDS (middle) in our Hycean K2-18 b model with modern Earth S$_{\textrm{org}}$ flux, computed by VULCAN 2D. The substellar point is located at 180$^{\circ}$ longitude. The right panel depicts the vertical distribution of several species across the equatorial region, where different shades correspond to different longitudinal locations. The vertical profiles of DMS excluding zonal transport are shown in red dotted (morning terminator, 73$^\circ$ westward sub-stellar point) and dashed-dotted (evening terminator, 73$^\circ$ eastward sub-stellar point) lines for comparison.}\label{fig:2D}
\end{figure}

\subsection{2D: DMS is homogenized by global circulation}
The temperature and wind structures simulated by our Hycean K2-18 b GCM can be found in Figure \ref{fig:gcm}. The circulation exhibits an eastward zonal jet in the troposphere and day-night thermally driven cells in the stratosphere. In the upper atmosphere above 10$^{-4}$ bar, the flow transitions to retrograde, likely due to vertical momentum transport \citep{Tsai2014}. The mean tropospheric zonal wind below 0.01 bar is about 20 m/s, consistent with the jet speed estimated from wave balance \citep{Hammond2020}. This zonal wind speed translates to a horizontal transport timescale of $\sim$ 10 days.

Figure \ref{fig:2D} illustrates the equatorial distribution of DMS and DMDS for modern Earth S$_{\textrm{org}}$ production. DMS exhibits a rather uniform abundance below 10$^{-4}$ bar. As the zonal winds change to retrograde in the upper atmosphere, DMS extends to higher altitudes around the evening terminator due to the westward transport from the nightside. On the other hand, DMDS is photochemically produced by \ce{CH3SH} via
\begin{eqnarray}
\begin{aligned}
\ce{CH3SH &->[h\nu] CH3S + CH3}\\
\ce{CH3S + CH3S &-> DMDS (CH3S2CH3)}
\end{aligned}
\label{re:CH3SH-DMDS}
\end{eqnarray}
in the stratosphere between 10$^{-2}$--10$^{-4}$ bar and follows the thermally driven cells. This relatively uniform distribution of DMS can be understood by comparing relevant timescales. The photochemical lifetime of DMS on the dayside ($\tau_{DMS}^{day}$) is determined by $\tau_{\textrm{DMS}}^{\textrm{day}} = 1/k_{\textrm{DMS}}$
, where $k_{\textrm{DMS}}$ is its photolysis rate. $\tau_{DMS}^{day}$ is $\sim$ 10$^8$--10$^9$ s in the troposphere below 10$^{-2}$ bar. On the nightside, the replenish timescale ($\tau_{DMS}^{night}$) is given by $\sim$ $H / K_{\textrm{zz}}^2$ $\sim$ 10$^8$ s. Given that $\tau_{DMS}^{day}$ and $\tau_{DMS}^{night}$ are comparable to each other and considerably longer than the horizontal transport timescale of 
$\sim$ 10$^6$ s, the S$_{\textrm{org}}$ species on the nightside is still constrained by the photochemical processes on the dayside through global transport. We note that our 2D model already assumes an optimistic scenario with a uniform S$_{\textrm{org}}$ flux across the nightside. In reality, the nightside S$_{\textrm{org}}$ flux might be much lower due to the lack of stellar energy input to power a (hypothetical) photosynthetic biosphere. Overall, the DMS abundance predicted by our 2D photochemical model is broadly consistent with that from 1D model, as indicated by the right panel in Figure \ref{fig:2D}. 




\begin{figure}[ht!]
\includegraphics[width=0.5\textwidth]{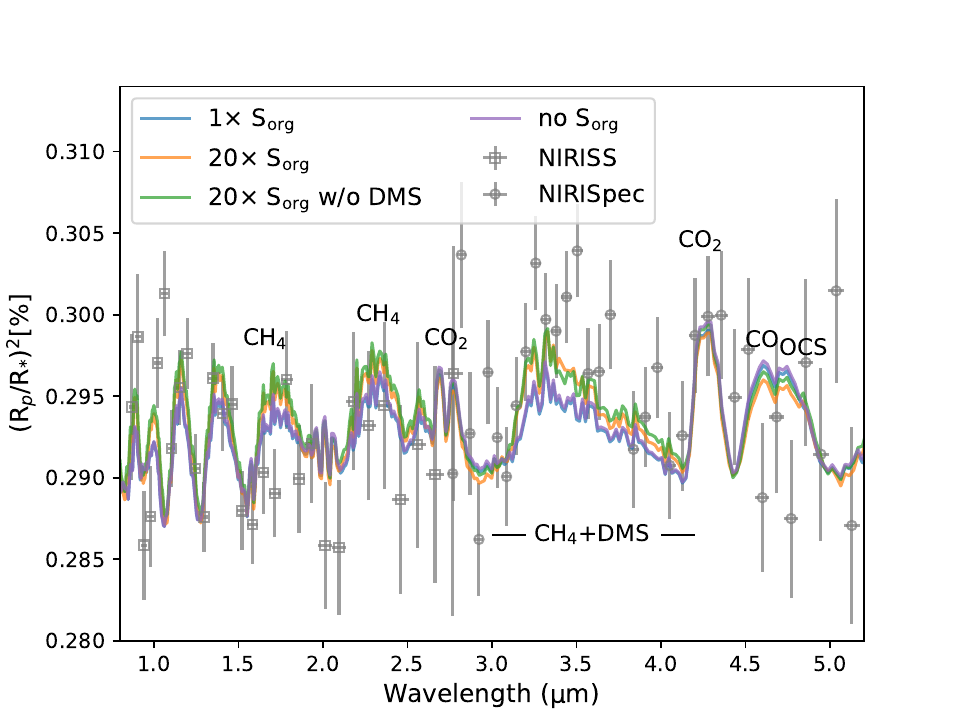}
\includegraphics[width=0.5\textwidth]{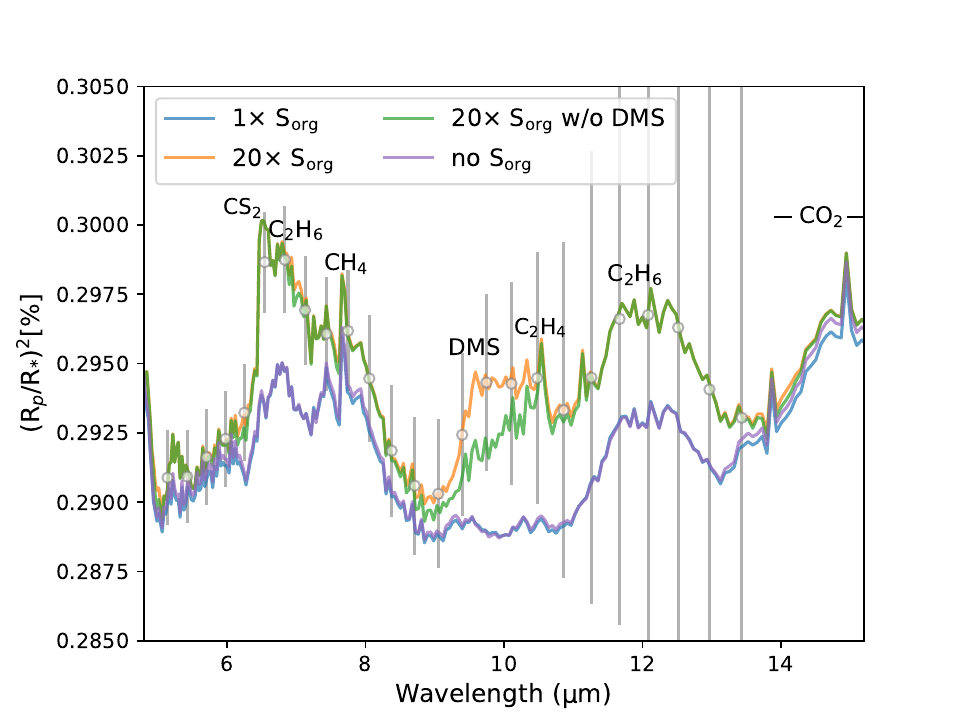}
\caption{Synthetic transmission spectra of Hycean K2-18 b 1D models for 1$\times$ and 20$\times$ modern Earth biological sulfur flux. The JWST NIRISS/SOSS and NIRSpec/G395H data from \citep{Madhusudhan2023} are plotted in the left panel. The simulated noise by Pandexo \citep{Batalha2017} with the JWST/MIRI LRS for 5 transits without random noise is plotted in the right panel (gray error bars). The 20$\times$ S$_{\textrm{org}}$ spectrum but with DMS opacity removed is shown for comparison. The ``no S$_{\textrm{org}}$'' case includes a flux for \ce{CH4} equivalent to Modern Earth's non-anthropogenic emission \citep{Shawn2011}, but omits fluxes for all biogenic sulfur gases. The corresponding abundance profiles used to produce the spectra can be found in Figure \ref{fig:1x-20x-profiles}.}\label{fig:spectra}
\end{figure}




\subsection{Transmission spectra}
We simulated synthetic transmission spectra of our cloud-/haze-free Hycean K2-18 b models using PLATON \citep{Zhang2019,Zhang2020}, including opacity sources of \ce{CH4}, \ce{CO},\ce{CO2}, \ce{C2H2}, \ce{H2O},\ce{HCN}, \ce{NH3}, \ce{O2}, \ce{NO}, \ce{C2H4}, \ce{C2H6}, \ce{H2CO}, \ce{N2}, \ce{NO2}, \ce{H2S}, \ce{COS}, \ce{SH}, \ce{SO2}, (CIA) of \ce{H2}–\ce{H2} and \ce{H2}–He. We further include cross-sections of \ce{CH3SH}, DMS, and DMS from the Pacific Northwest National Laboratories (PNNL) database \citep{PNNL2004}, which are limited to 1 bar pressure and room temperature. 

The transmission spectra for none, 1$\times$ and 20$\times$ modern Earth S$_{\textrm{org}}$ flux are compared in Figure \ref{fig:spectra}. For the JWST NIRISS and NIRSpec coverage below 5 $\mu m$, all of our Hycean K2-18 b models with methanogenic flux produce \ce{CH4} absorption features between 1 and 4 $\mu m$ that are broadly consistent with the JWST data. However, it is challenging to distinguish the DMS absorption contribution within the NIRSpec coverage, even with 20$\times$ S$_{\textrm{org}}$ flux. The signal of DMS around 3.5 $\mu m$ for 20$\times$ S$_{\textrm{org}}$ flux does not exceed 10 ppm, smaller than NIRSpec's precision of 30--50 ppm at R $\sim$ 50 \citep{Madhusudhan2023}. The most promising diagnostics window of DMS is in the mid-infrared, as indicated in the right panel of Figure \ref{fig:spectra}. For the 20$\times$ S$_{\textrm{org}}$ case, DMS and the photochemical byproduct \ce{C2H4} contribute to an absorption of about 50 ppm at 9--11 $\mu m$, in addition to the \ce{C2H6} feature at 7$\mu m$. The 7--$\mu m$ feature of \ce{C2H6} also overlaps with \ce{CS2}, making it difficult to differentiate the contribution of \ce{CS2}. Our transmission spectra reveal no discernible \ce{H2S} spectral signatures. While \ce{H2S} can reach above ppm levels with outgassing and photochemical sources (e.g. pathway (\ref{re:CH3SH-H2S})), its broad opacities that overlap with several species make it challenging to detect.  

    Since ethane (\ce{C2H6}) is efficiently produced from \ce{CH4} and has strong mid-infrared absorption bands, \ce{C2H6} shows a prominent feature at 7 and 11--13 $\mu m$ with or without the presence of S$_{\textrm{org}}$ flux. By constraining [\ce{CH4}]/[\ce{C2H6}], one can potentially estimate the relative contribution of methanogenic \ce{CH4} flux relative to S$_{\textrm{org}}$ flux \citep{Shawn2011}. Since our model likely overpredicts \ce{C6H6} in the gas phase due to the lack of conversion to higher-order hydrocarbons and particles, we did not include \ce{C6H6} opacities when generating the spectra in Figure \ref{fig:spectra}. However, it is interesting to keep in mind that the presence ($\gtrsim$ 50 ppm) of \ce{C6H6} would exhibit detectable features within NIRSpec wavelength range, in addition to \ce{CH4} and \ce{CO2} (see Figure \ref{fig:c6h6_spectra}). For biogenic flux $\gtrsim$ 30$\times$ Earth's S$_{\textrm{org}}$ flux, the atmospheric mean molecular weight exceeds 8 amu in addition to the expected aerosols that would flatten transmission spectral features, making it immediately distinguishable for $\lesssim$ 20$\times$ cases. We summarize the identifiable biosignature features with 1$\times$--20$\times$ S$_{\textrm{org}}$ flux in Table 1.  

\newcommand{\cmark}{\ding{51}}
\newcommand{\xmark}{\ding{55}}%
\definecolor{bleudefrance}{rgb}{0.19, 0.55, 0.91}
\begin{table}[!h]
\centering
\caption{Detectability of potential biosignature gases with corresponding wavelength ranges on Hycean K2-18 b based on our models with 1$\times$--20$\times$ S$_{\textrm{org}}$ flux. {\color{orange} \cmark}  denotes detectable ($\gtrsim$ 50 ppm) and {\color{bleudefrance} \xmark}  denotes nondetectable ($\lesssim$ 10 ppm), whereas {\bf ?} means challenging, with shallow features ($\gtrsim$ 20 ppm but $\lesssim$ 50 ppm) and only present under enhanced 20$\times$ S$_{\textrm{org}}$ flux.}
\begin{tabular}{c c c l}  \hline \hline
Species & 1--5 $\mu m$ & 5--15 $\mu m$ \\ \hline
\ce{CH4} & {\color{orange} \cmark} & {\color{orange} \cmark} \\
\ce{CO2} & {\color{orange} \cmark} & {\color{orange} \cmark} \\
\ce{H2S} & {\color{bleudefrance}\xmark} & {\color{bleudefrance}\xmark} \\
\ce{CH3SH} & {\color{bleudefrance}\xmark} & {\color{bleudefrance}\xmark} \\
\ce{CS2} & {\color{bleudefrance}\xmark} & {\bf ?} \\
\ce{DMS} & {\color{bleudefrance}\xmark} & {\bf ?} \\
\ce{DMDS} & {\color{bleudefrance}\xmark} & {\color{bleudefrance}\xmark} \\
\ce{C2H4} & {\color{bleudefrance}\xmark} & {\bf ?} \\
\ce{C2H6} & {\color{bleudefrance}\xmark} & {\color{orange} \cmark} \\
\hline
\end{tabular}
\label{tab:vulcan_parameters}
\end{table}

\section{Discussion}
Since K2-18 b requires reflective clouds and hazes to sustain water oceans, we explore the potential UV shielding effects due to clouds and hazes in the bottom panel of Figure \ref{fig:spectra_tau}. Organic sulfur species only start to build up when the stellar UV is reduced to less than 20$\%$. For K2-18 b specifically, \cite{Leconte2024} highlighted the observed methane features deeper than $\sim$ 100 ppm (their Fig. 9) are not consistent with highly reflective clouds. We argue that our photochemical results should remain robust to UV attenuation by clouds and hazes, unless muted spectral features indicate scattering by high-altitude aerosols can significantly reduce stellar UV flux.

The predicted requirement of 20$\times$ S$_{\textrm{org}}$ flux to be potentially detectable is equivalent to about 0.004 g m$^{-2}$ based on the DMS lab production rate \citep{Seager2013}, which is within the plausible range of surface biomass density. Our predictions for the biogenic sulfur gases in Hycean atmospheres are broadly consistent with previous studies of the Archean Earth \cite{Shawn2011} and Archean-like atmospheres on Trappist-1 planets \citep{Meadows2023}. While \cite{Meadows2023} concluded \ce{CH3SH} to be the most prominent S$_{\textrm{org}}$ feature in \ce{N2}-dominated atmospheres, we find efficient \ce{CH3SH} conversion into \ce{H2S} in a \ce{H2} atmosphere. Consequently, DMS appears to constitute the most detectable organic sulfur molecule in the mid-infrared for Hycean worlds.  

On modern Earth, the oxidation of DMS might participate in the sulfuric cloud formation \citep{Charlson1987} and affect climate feedback. It has been suggested that this biologically driven feedbacks could help stabilize the Earth's climate \citep{Lovelock1989}, although the precise pathways and quantified contributions from DMS to aerosols still remain elusive. Similarly, on a Hycean world with highly elevated S$_{\textrm{org}}$ flux ($\gtrsim$30$\times$ modern Earth's value, see Figure 1.), hydrocarbon and elemental sulfur hazes (\ce{S8}) are expected to form. We speculate a similar negative feedback loop can take place in a Hycean world, where photochemical hazes derived from organic sulfur flux can increase albedo and prevent a runaway state.

Lastly, we reiterate that constructing complete line lists for DMS, DMDS, and \ce{CH3SH} to correctly account for the pressure-broadening effects would improve the interpretation of spectral data. Future work extending the study encompassing the diversity of halomethanes, such as methyl chloride and methyl bromide \citep[e.g., ][]{Leung2022},  will expand our understanding of the biosignature gases on sub-Neptune waterworlds. 

\section{Conclusion}
We find a biogenic sulfur flux of approximately 20 times higher than modern Earth's is required for DMS to reach detectable levels (i.e. above $\sim$ppm) on a K2-18b-like Hycean world. At the terminators, our 2D photochemical model shows only minor abundance enhancements in DMS compared to 1D model predictions. On the other hand, JWST/MIRI could detect \ce{C2H6} around 7 $\mu m$ and the broad feature at 9--13 $\mu m$ contributed from DMS, \ce{C2H4}, and \ce{C2H6} with 20$\times$ S$_{\textrm{org}}$ flux, based on our fiducial Hycean model. However, identifying DMS signatures within NIRSpec's wavelength range is demonstrated to be challenging, despite DMS appearing as the most promising biogenic sulfur signature directly from the S$_{\textrm{org}}$ source. The moderate threshold for biological production suggests that the search for biogenic sulfur gases as one class of potential biosignature is plausible for Hycean worlds.

\begin{acknowledgments} 
The authors thank Michaela Leung for the useful comments on an early draft of this article. The authors also thank Stephen Klippenstein and Julie I. Moses for the illuminating discussion on \ce{H2CO} kinetics and for sharing the ab initio calculations. S.-M.T. thanks Eva-Maria Ahrer, Guangwei Fu, Jake Taylor, and Lili Alderson for helpful discussions on data resampling and feature identification. S.-M.T. and E.W.S. acknowledge support from National Aeronautics and Space Administration (NASA) through Exobiology Grant No. 80NSSC20K1437 and Interdisciplinary Consortia for Astrobiology Research (ICAR) Grant Nos. 80NSSC23K1399, 80NSSC21K0905, and 80NSSC23K1398. H.I. is funded by the European Union (ERC, DIVERSE, 101087755 and ERC, EXOCONDENSE, 740963). Views and opinions expressed are however those of the author(s) only and do not necessarily reflect those of the European Union or the European Research Council Executive Agency. Neither the European Union nor the granting authority can be held responsible for them. N.F.W. was supported by the NASA Postdoctoral Program.
\end{acknowledgments}

\software{Exo-FMS \citep{Lee2020,Lee2021},  
          VULCAN \citep{Tsai2017,Tsai2021},
          Numpy \citep{Walt2011}
          }   

\newpage          
\appendix

\renewcommand{\thefigure}{A\arabic{figure}}
\renewcommand{\theHfigure}{A\arabic{figure}}
\renewcommand{\thetable}{A\arabic{table}}
\renewcommand{\theHtable}{A\arabic{table}}

\setcounter{figure}{0}
\setcounter{table}{0}

\section{Model input parameters}\label{app:parameters}
Table \ref{tab:planet_parameters} lists the input parameters for Hycean K2-18 b with the HELIOS radiative transfer model and the ExoFMS general circulation model. HELIOS includes \ce{H2O}, \ce{CH4}, CO, \ce{CO2}, \ce{NH3}, HCN, \ce{C2H2}, collision-induced absorption (CIA) of \ce{H2}–\ce{H2} and \ce{H2}–He as opacity sources. We ignore the \ce{H2O} self continuum absorption and likely underestimate the tropospheric temperature. ExoFMS uses double-gray opacities for its radiative transfer. The optical depth in the infrared ($\tau_{\text{IR}}$) and shortwave ($\tau_{\text{SW}}$) bands is given by:
\begin{equation}
d \tau_{\text{i}} = \qty(\kappa_{\text{i,d}}(1-q) +\kappa_{\text{i,w}}q)\qty(f_{\text{i}} + 2(1 - f_{\text{i}}))\qty(\frac{p}{p_s}))\frac{dp}{g},
\end{equation}
where the $i$ subscript represents either the SW or IR band, $p$ is the pressure and $p_s$ the surface pressure. The opacities $\kappa_{i,w}$ and $\kappa_{i,d}$ represent the opacities due to the water vapour component and dry component (all other gases) of the atmosphere respectively. The factor $f_i$ accounts for pressure dependence of opacity, due to collision-induced absorption and pressure broadening effects. To reproduce the inversion in the HELIOS calculations, a second band was added in the SW part of the spectrum, receiving a fraction $\alpha$ of the total incoming flux. The optical depth in this band was given by:
\begin{equation}
    \tau_{SW,2} = \frac{\kappa_{\text{SW,2}} p}{g}.
\end{equation}
The water mass concentration, $q$, used in this calculation is taken from the 1D HELIOS calculation and remains constant throughout the simulation (though in reality this should be spatially inhomogeneous and linked to the local saturation vapour pressure at a given temperature and pressure). The value of $\kappa_{IR,w}$ was taken to match the runaway greenhouse OLR of $\approx 280$ Wm$^{-2}$ \citep[as in][]{Innes2023}. The other values were tuned to match the HELIOS temperature-pressure profile. The parameters used are shown in Table~\ref{tab:rad_parameters}.

\renewcommand*{\thefootnote}{\fnsymbol{footnote}}
\begin{table}[!h]
\centering
\caption{Planetary parameters for Hycean K2-18 b.}
\begin{tabular}{c c c l}  \hline \hline
Symbol & Value  & Unit & Description \\ \hline
R$_{\rm p}$ & 1.66 $\times$ 10$^{7}$  & m & Planetary radius\\
a & 0.1591 & AU & Orbital distance\\
T$_{\rm irr}$ & 1370 & K & Irradiation temperature \\
T$_{\rm int}$ & 0 & K & Internal temperature \\
g & 12.43  & m s$^{-2}$ & Gravitational acceleration \\
c$_{\rm p}$ & 2216  &  J K$^{-1}$ kg$^{-1}$ & Specific heat capacity \\
$\Omega_{\rm p}$ & 2.2137 $\times$ 10$^{-6}$ & rad s$^{-1}$ & Planetary rotation rate \\
\hline
\end{tabular}
\label{tab:planet_parameters}

\end{table}

\begin{table}[!h]
\centering
\caption{Radiative parameters for the ExoFMS double-gray radiation scheme.}
\begin{tabular}{c c c l}  \hline \hline
Symbol & Value  & Unit & Description \\ \hline
$\kappa_{IR,d}$ & \num{1.15e-3}& \si{\m\squared\per\kg}& IR dry opacity\\
$\kappa_{\text{IR,w}}$ &\num{1.e-2} & \si{\m\squared\per\kg} & IR water opacity\\
$\kappa_{SW,d}$ & \num{2.e-5} & \si{\m\squared\per\kg} & SW dry opacity\\
$\kappa_{\text{SW,w}}$ & \num{4.e-4} & \si{\m\squared\per\kg} & SW water opacity \\
$\kappa_{\text{SW,2}}$ & \num{5.e-2} & \si{\m\squared\per\kg} & SW second band opacity\\
$\alpha$ & \num{1.e-2} & Dimensionless & Fraction of instellation in SW second band\\
$f_{\text{IR}}$ & 0.9 & Dimensionless & Controls pressure dependence of $\tau_{\text{IR}}$\\
$f_{\text{SW}}$ & 0.2 & Dimensionless & Controls pressure dependence of $\tau_{\text{SW}}$\\
\hline
\end{tabular}
\label{tab:rad_parameters}
\end{table}

\begin{table}[!h]
\centering
\caption{Lower boundary conditions for photochemical modeling of inhabited Hycean K2-18 b.}
\begin{tabular}{c c c} \hline \hline
Species & Surface Emission (molecules cm$^{-2}$ s$^{-1}$)  & Vdep (cm s$^{-1}$) \\ \hline
\ce{CH4}$^\text{a}$ & 7$\times$ 10$^{10}$ & 0\\
\ce{CH3SH}$^\text{a}$ & 8.3$\times$ 10$^{8}$ & 0\\
\ce{DMS}$^\text{a}$ & 4.2$\times$ 10$^{9}$ & 0$^\text{d}$\\
\ce{CS2}$^\text{a}$ & 1.4$\times$ 10$^{7}$ & 0\\
\ce{H2S}$^\text{b}$ & 2$\times$ 10$^{8}$ & 0.015\\
\ce{SO2}$^\text{b}$ & 9$\times$ 10$^{9}$ & 1\\
\ce{COS}$^\text{b}$ & 5.4$\times$ 10$^{7}$ & 0.003\\
\ce{H2O2}$^\text{c}$ & 0 & 1\\
\ce{CH3S}$^\text{a}$ & 0 & 0.01\\
\ce{HSO}$^\text{a}$ & 0 & 1\\
\ce{S}$^\text{a}$ & 0 & 1\\
\ce{SO}$^\text{a}$ & 0 & 0.0003\\
\hline
\multicolumn{3}{p{0.5\linewidth}}{
\raggedright
$^\text{a}$ \cite{Shawn2011}\\
$^\text{b}$ \cite{Seinfeld2016}\\
$^\text{c}$ \cite{Hauglustaine1994}\\
$^\text{d}$ For the scenario that the surface layer of the ocean is saturated. The measured deposition velocity of DMS on land is 0.064--0.28 cm s$^{-1}$ \citep{Judeikis1977}.\\
}
\end{tabular}
\label{tab:vulcan_parameters}
\end{table}
\begin{figure}[ht!]
\includegraphics[width=0.99\textwidth]{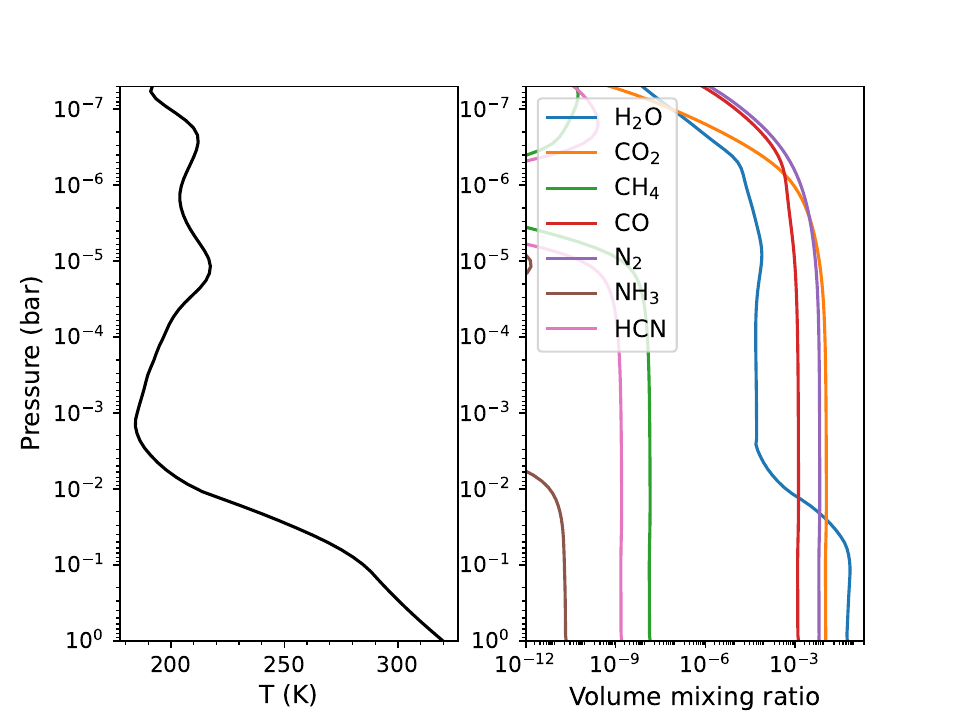}
\caption{The self-consistent temperature (left) and composition (right) profiles generated by iterative runs between HELIOS and VULCAN. Sulfur kinetics is not included in the VULCAN runs here for simplicity.}
\label{fig:no-S-TP}
\end{figure}

\begin{figure}[ht!]
\includegraphics[width=0.52\columnwidth]{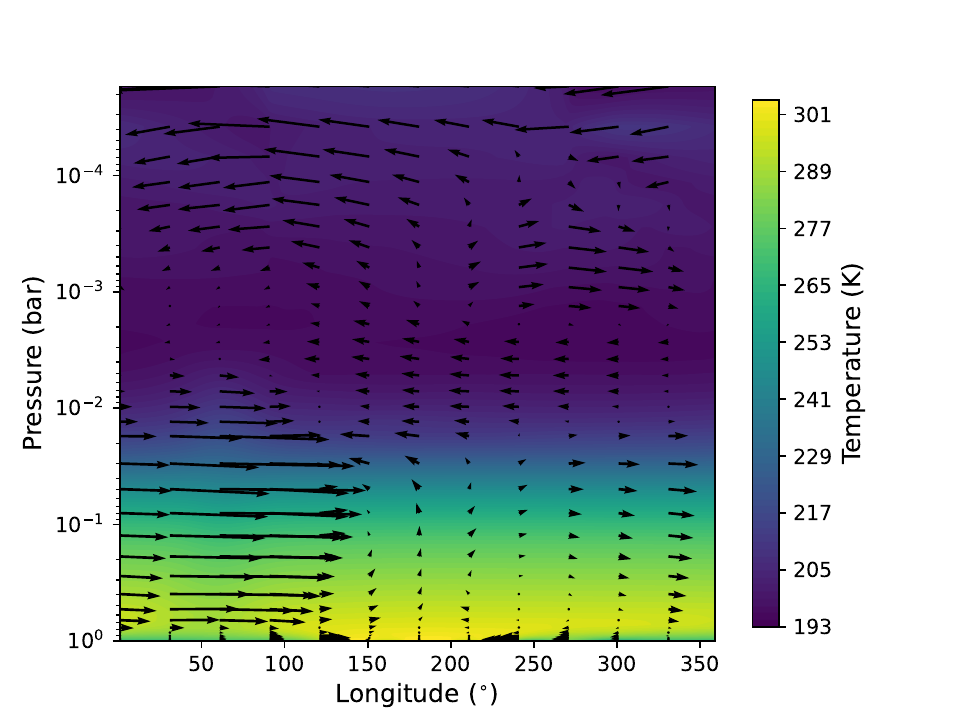}
\includegraphics[width=0.49\columnwidth]{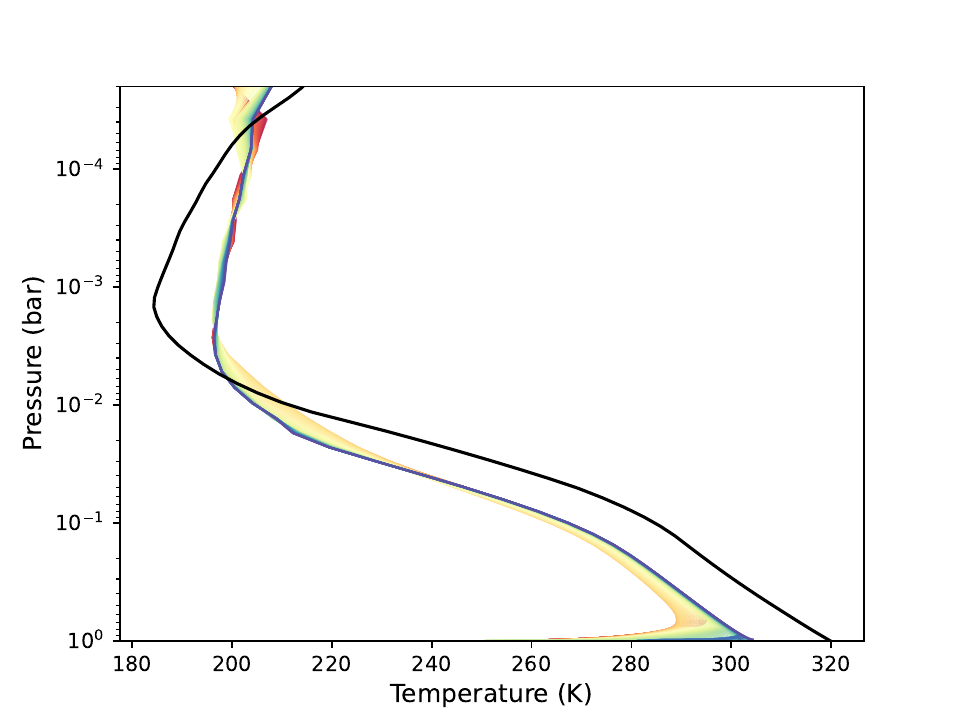}
\includegraphics[width=0.52\columnwidth]{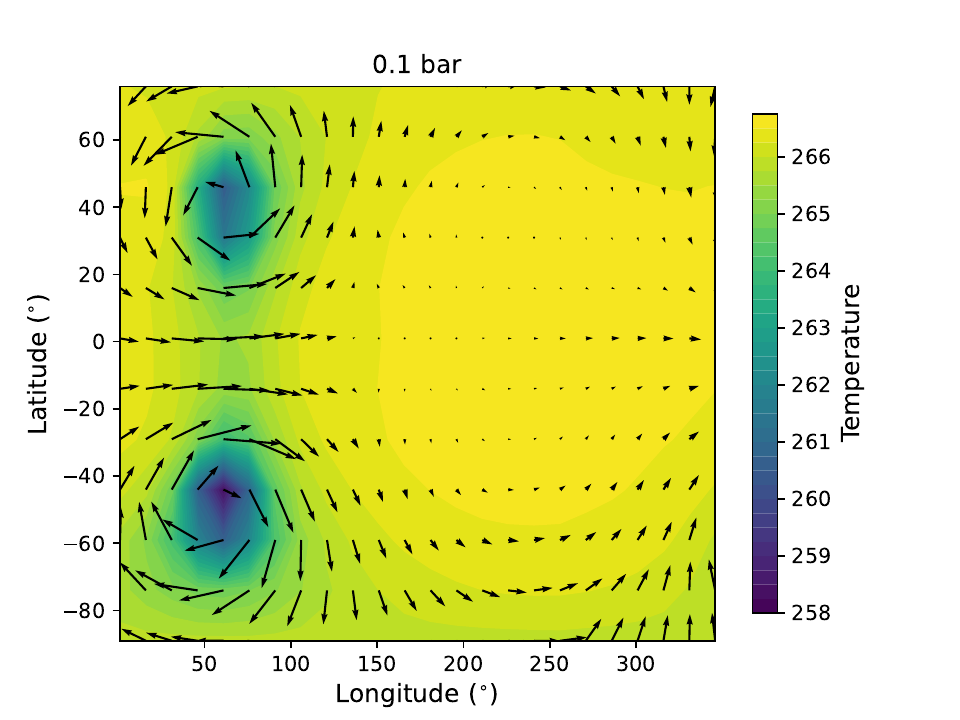}
\includegraphics[width=0.52\columnwidth]{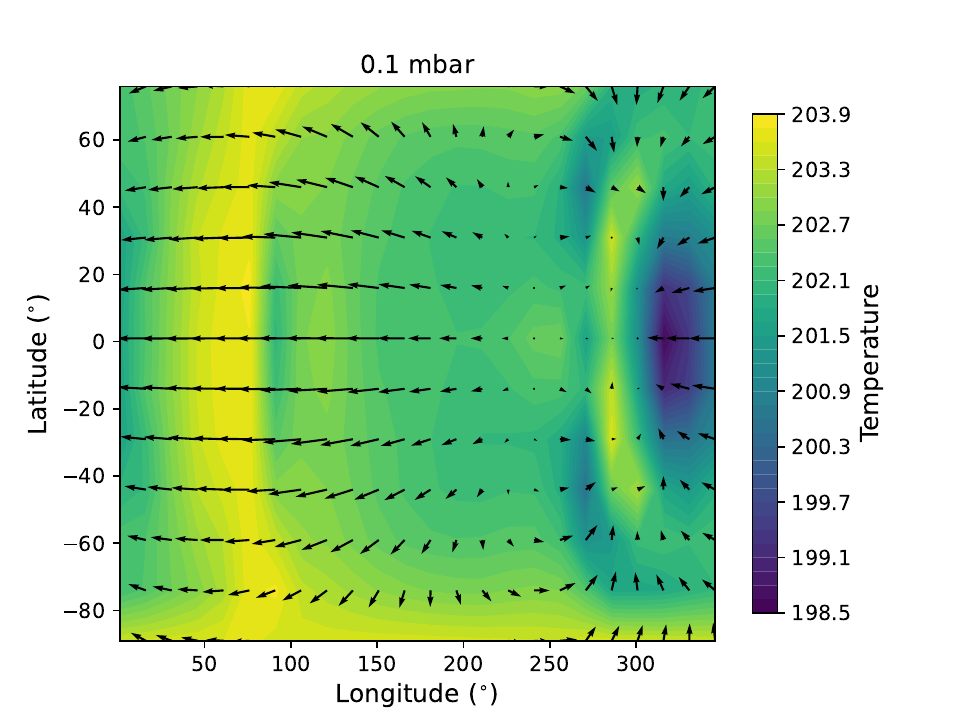}
\caption{The top left panel shows the 
temperatures (color scale) and winds (arrows) on the equatorial plane averaged across $\pm$30$^{\circ}$ latitudes from our K2-18 b GCM (substellar point located at 180$^{\circ}$ longitude). The top right panel displays the vertical temperature profiles around the equator corresponding to different longitudinal locations, where the temperature profile from the 1D radiative transfer calculation (HELIOS) is plotted in black for comparison. The bottom two panels show the temperatures and winds at 0.1 bar and 0.1 mbar level.}
\label{fig:gcm} 
\end{figure}

\begin{figure}[ht!]
\includegraphics[width=0.5\columnwidth]{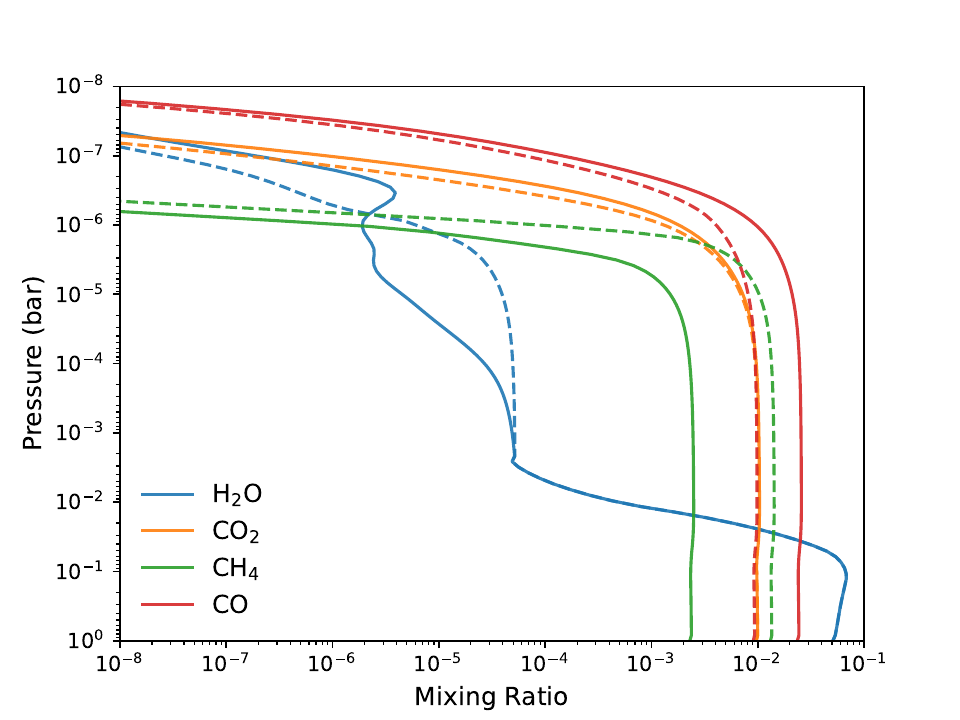}
\includegraphics[width=0.5\columnwidth]{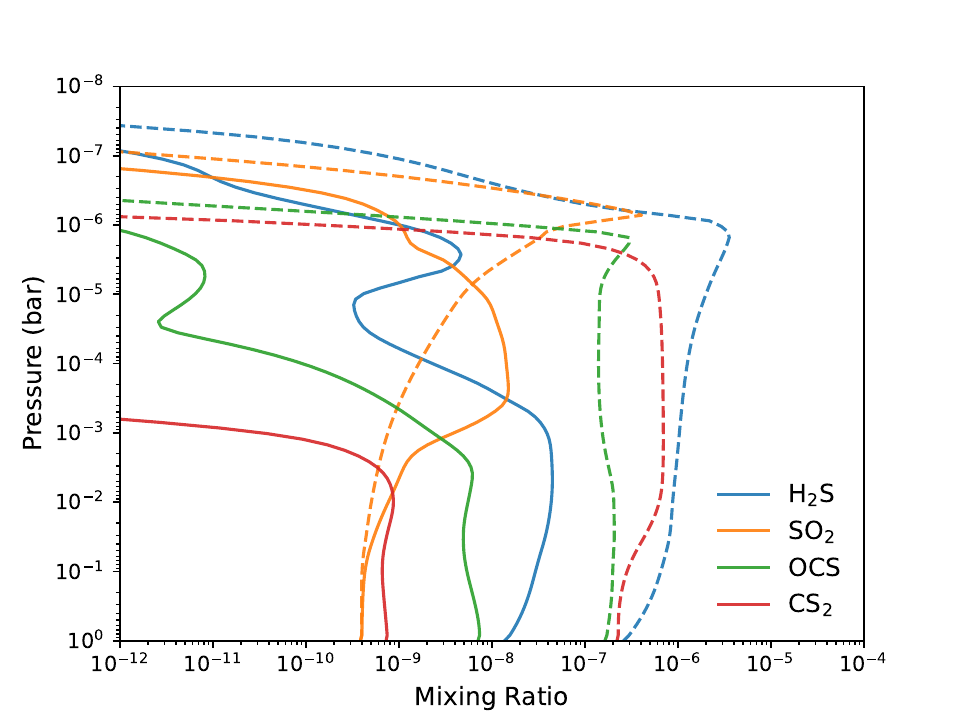}
\includegraphics[width=0.5\columnwidth]{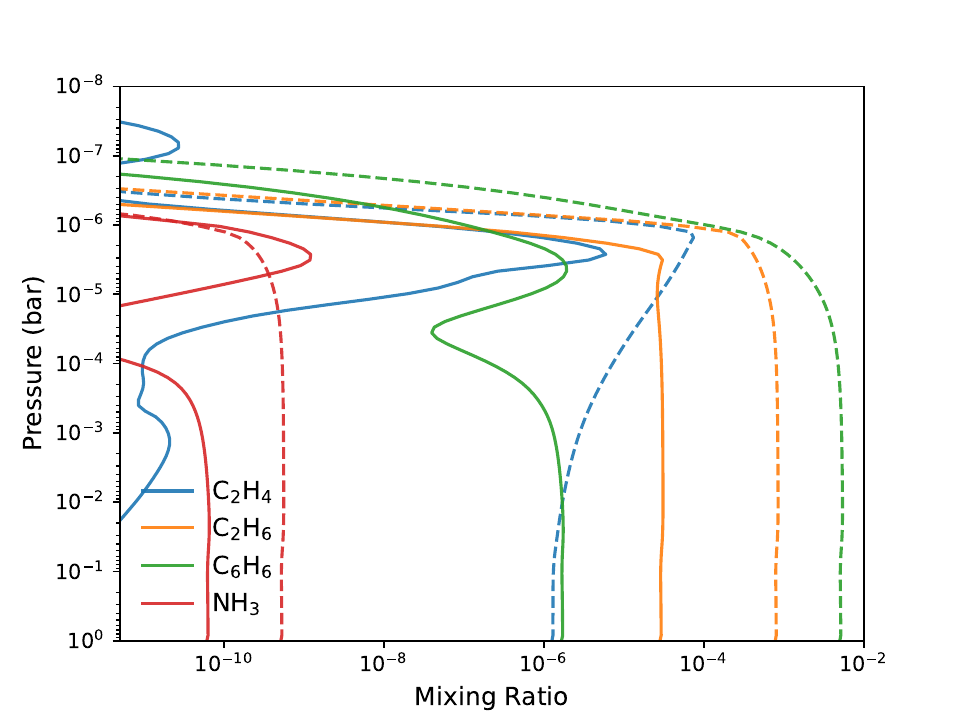}
\includegraphics[width=0.5\columnwidth]{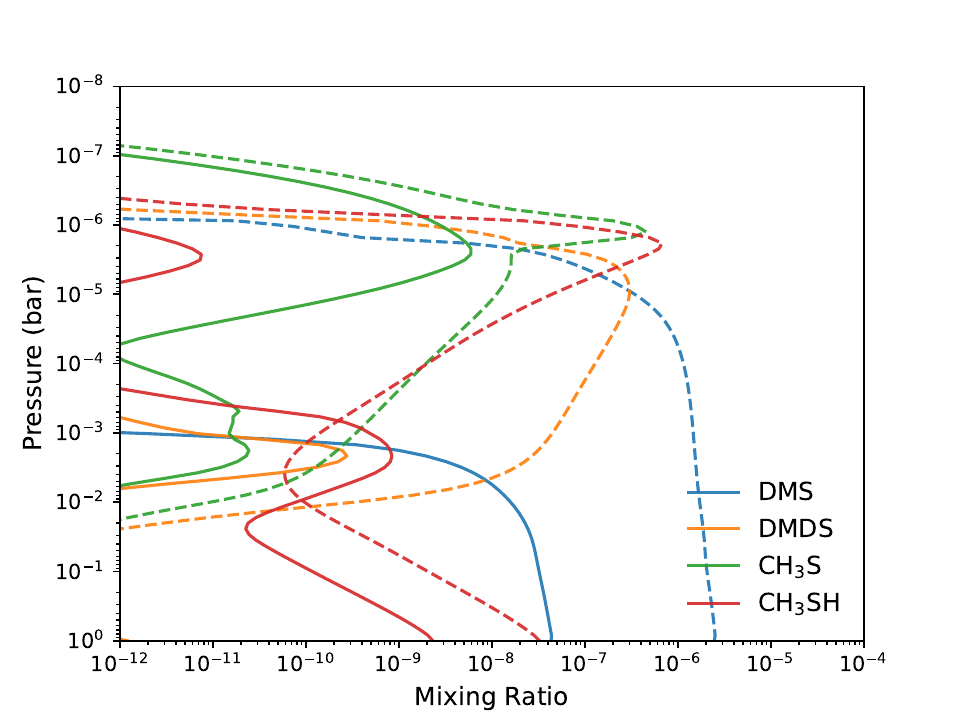}
\caption{The abundance profiles of the Hycean K2-18 b 1D models with 1$\times$ (solid) and 20$\times$ (dashed) modern Earth values. 
}
\label{fig:1x-20x-profiles} 
\end{figure}

\begin{figure}[ht!]
\includegraphics[width=0.5\columnwidth]{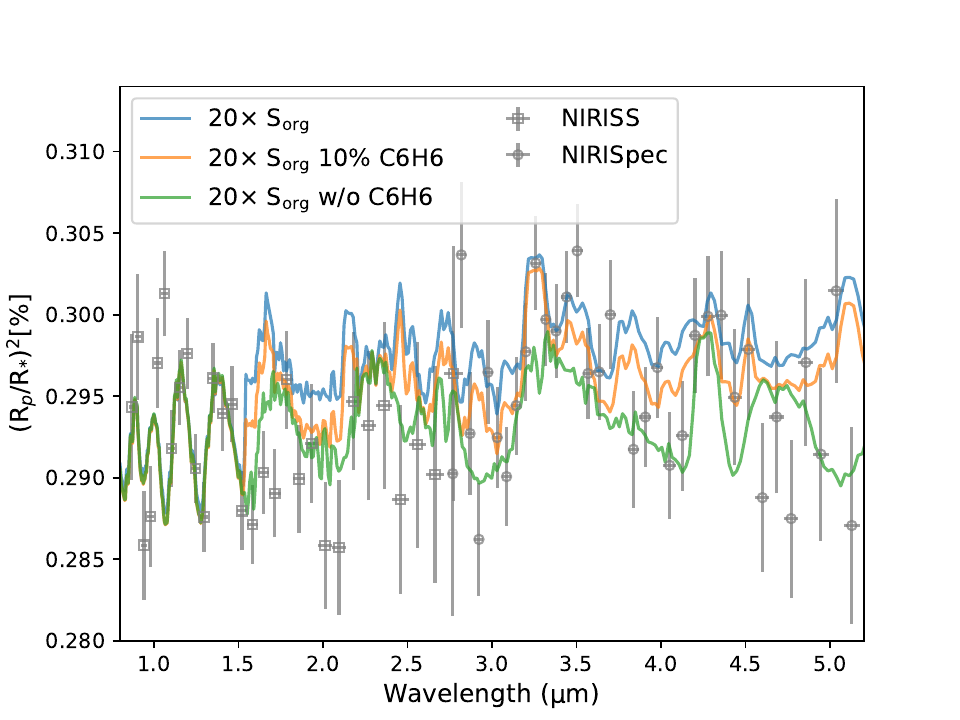}
\caption{Synthetic transmission spectra of Hycean K2-18 b similar to Figure \ref{fig:spectra}, but including \ce{C6H6} absorption for 20$\times$ S$_{\textrm{org}}$ flux. The spectra without \ce{C6H6} opacity and with 10$\%$ of \ce{C6H6}, representing depletion from haze formation, are shown for comparison.  
}
\label{fig:c6h6_spectra} 
\end{figure}

\section{Updated H$_2$CO reaction rates}\label{app:h2co}
Formaldehyde (\ce{H2CO}) is one of the main intermediate species in \ce{CO2}-CO-\ce{CH4} interconversion. Forming and breaking the double bond between C and O in \ce{H2CO} may facilitate an important step that controls the overall timescale \citep{Tsai2018}. There are two three-body reactions involving \ce{H2CO} that we adopt the thermal dissociation rate coefficients due to the lack of low-pressure rate limits for recombination:
\begin{equation}
\ce{CH2OH ->[M] H + H2CO}\label{H2CO_1}
\end{equation}
and
\begin{equation}
\ce{CH3O ->[M] H + H2CO}\label{H2CO_2}
\end{equation}
The recombination rates are then reversed by thermochemical data. In the default chemical networks in VULCAN (e.g. \href{https://github.com/exoclime/VULCAN/blob/master/thermo/NCHO_photo_network.txt}{NCHO\_photo\_network}), the adopted rate coefficients either underestimate the activation energy or are only valid at high temperatures. As a result, these reversed rates tend to overpredict the \ce{CO2}--\ce{CH4} conversion efficiency around room temperatures. In this study, we have adopted 
the low-pressure rate coefficients for \ref{H2CO_1} and \ref{H2CO_2} from \cite{Tsang1986} and \cite{Tsang1987}, which are more consistent with the ab initio calculations conducted by \cite{Klippenstein2023}. The original and updated rate coefficients are summarized in Table \ref{tab:H2CO}. This update of \ce{H2CO} kinetics significantly lowers the \ce{CH4} abundance in our lifeless case (without methanogenic flux; Figure \ref{fig:no-S-TP})) from $\sim$10$^{-5}$ to $\sim$10$^{-8}$, which is consistent with the calculations in \cite{Wogan2024}.

\begin{table}[!h]
\centering
\caption{}
\begin{tabularx}{.9\linewidth}{p{0.27\linewidth} p{0.34\linewidth} p{0.25\linewidth} }
Reactions & Rate Coefficients$^\text{a}$ & Reference\\ \hline
\multicolumn{3}{c}{Original} \\ \hline
\ce{CH2OH ->[M] H + H2CO} & $k_0 = 1.66 \times 10^{-10} \exp(-12630/T)$ \newline $k_{\infty} = 3 \times 10^{9} \exp(-14640/T)$ & \cite{Cribb1992} \newline \cite{Tsuboi1981}\\
\ce{CH3O ->[M] H + H2CO} & $k_0 = 9 \times 10^{-11} \exp(-6790/T)$ \newline 
    $k_{\infty} = 1.56 \times 10^{15} T^{-0.39} \exp(-13300/T)$ & \cite{Baulch1994} \newline \cite{Curran2006}\\  \hline
\multicolumn{3}{c}{Updated} \\ \hline
\ce{CH2OH ->[M] H + H2CO} & $k_0 = 7.48 \times 10^{1} T^{-2.5} \exp(-17200/T)$ \newline $k_{\infty} = 4.53 \times 10^{34} T^{-7.11} \exp(-22200/T)$ & \cite{Tsang1987} \newline \cite{Xu2015}\\
\ce{CH3O ->[M] H + H2CO} & $k_0 = 6.51 \times 10^{13} T^{-6.65} \exp(-16700/T)$ \newline
    $k_{\infty} = 3.17 \times 10^{24} T^{-4.25} \exp(-13100/T)$ & \cite{Tsang1986} \newline \cite{Xu2015}\\  \hline
\multicolumn{3}{p{0.9\linewidth}}{
      $^\text{a}$In low pressure rate coefficient, $k_0$ (cm$^6$ molecules$^{-2}$ s$^{-1}$), and high pressure rate coefficient, $k_{\infty}$ (cm$^3$ molecules$^{-1}$ s$^{-1}$).}
\end{tabularx}
\label{tab:H2CO}
\end{table}

\bibliographystyle{aasjournal}
\bibliography{master_bib.bib}

\begin{thebibliography}{}
\expandafter\ifx\csname natexlab\endcsname\relax\def\natexlab#1{#1}\fi
\providecommand{\url}[1]{\href{#1}{#1}}
\providecommand{\dodoi}[1]{doi:~\href{http://doi.org/#1}{\nolinkurl{#1}}}
\providecommand{\doeprint}[1]{\href{http://ascl.net/#1}{\nolinkurl{http://ascl.net/#1}}}
\providecommand{\doarXiv}[1]{\href{https://arxiv.org/abs/#1}{\nolinkurl{https://arxiv.org/abs/#1}}}

\bibitem[{{Arney} {et~al.}(2018){Arney}, {Domagal-Goldman}, \& {Meadows}}]{Arney2018}
{Arney}, G., {Domagal-Goldman}, S.~D., \& {Meadows}, V.~S. 2018, Astrobiology, 18, 311, \dodoi{10.1089/ast.2017.1666}

\bibitem[{{Batalha} {et~al.}(2017){Batalha}, {Mandell}, {Pontoppidan}, {Stevenson}, {Lewis}, {Kalirai}, {Earl}, {Greene}, {Albert}, \& {Nielsen}}]{Batalha2017}
{Batalha}, N.~E., {Mandell}, A., {Pontoppidan}, K., {et~al.} 2017, \pasp, 129, 064501, \dodoi{10.1088/1538-3873/aa65b0}

\bibitem[{Baulch {et~al.}(1994)Baulch, Cobos, Cox, Frank, Hayman, Just, Kerr, Murrells, Pilling, Troe, Walker, \& Warnatz}]{Baulch1994}
Baulch, D.~L., Cobos, C.~J., Cox, R.~A., {et~al.} 1994, Journal of Physical and Chemical Reference Data, 23, 847, \dodoi{10.1063/1.555953}

\bibitem[{{Bean} {et~al.}(2021){Bean}, {Raymond}, \& {Owen}}]{Bean2021}
{Bean}, J.~L., {Raymond}, S.~N., \& {Owen}, J.~E. 2021, Journal of Geophysical Research (Planets), 126, e06639, \dodoi{10.1029/2020JE006639}

\bibitem[{{Benneke} {et~al.}(2019){Benneke}, {Wong}, {Piaulet}, {Knutson}, {Lothringer}, {Morley}, {Crossfield}, {Gao}, {Greene}, {Dressing}, {Dragomir}, {Howard}, {McCullough}, {Kempton}, {Fortney}, \& {Fraine}}]{Benneke2019}
{Benneke}, B., {Wong}, I., {Piaulet}, C., {et~al.} 2019, \apjl, 887, L14, \dodoi{10.3847/2041-8213/ab59dc}

\bibitem[{{Cala} {et~al.}(2023){Cala}, {Archer-Nicholls}, {Weber}, {Abraham}, {Griffiths}, {Jacob}, {Shin}, {Revell}, {Woodhouse}, \& {Archibald}}]{Cala2023}
{Cala}, B.~A., {Archer-Nicholls}, S., {Weber}, J., {et~al.} 2023, Atmospheric Chemistry \& Physics, 23, 14735, \dodoi{10.5194/acp-23-14735-2023}

\bibitem[{{Catling} \& {Zahnle}(2020)}]{Catling2020}
{Catling}, D.~C., \& {Zahnle}, K.~J. 2020, Science Advances, 6, eaax1420, \dodoi{10.1126/sciadv.aax1420}

\bibitem[{Charlson {et~al.}(1987)Charlson, Lovelock, Andreae, \& Warren}]{Charlson1987}
Charlson, R.~J., Lovelock, J.~E., Andreae, M.~O., \& Warren, S.~G. 1987, Nature, 326, 655, \dodoi{10.1038/326655a0}

\bibitem[{{Charnay} {et~al.}(2015){Charnay}, {Meadows}, \& {Leconte}}]{Charnay2015}
{Charnay}, B., {Meadows}, V., \& {Leconte}, J. 2015, \apj, 813, 15, \dodoi{10.1088/0004-637X/813/1/15}

\bibitem[{{Chen} {et~al.}(2018){Chen}, {Wolf}, {Kopparapu}, {Domagal-Goldman}, \& {Horton}}]{Chen2018}
{Chen}, H., {Wolf}, E.~T., {Kopparapu}, R., {Domagal-Goldman}, S., \& {Horton}, D.~E. 2018, \apjl, 868, L6, \dodoi{10.3847/2041-8213/aaedb2}

\bibitem[{Cribb {et~al.}(1992)Cribb, Dove, \& Yamazaki}]{Cribb1992}
Cribb, P.~H., Dove, J.~E., \& Yamazaki, S. 1992, Combustion and Flame, 88, 169, \dodoi{https://doi.org/10.1016/0010-2180(92)90050-Y}

\bibitem[{Curran(2006)}]{Curran2006}
Curran, H.~J. 2006, International Journal of Chemical Kinetics, 38, 250, \dodoi{https://doi.org/10.1002/kin.20153}

\bibitem[{{Domagal-Goldman} {et~al.}(2011){Domagal-Goldman}, {Meadows}, {Claire}, \& {Kasting}}]{Shawn2011}
{Domagal-Goldman}, S.~D., {Meadows}, V.~S., {Claire}, M.~W., \& {Kasting}, J.~F. 2011, Astrobiology, 11, 419, \dodoi{10.1089/ast.2010.0509}

\bibitem[{{Esparza-Borges} {et~al.}(2023){Esparza-Borges}, {L{\'o}pez-Morales}, {Adams Redai}, {Pall{\'e}}, {Kirk}, {Casasayas-Barris}, {Batalha}, {Rackham}, {Bean}, {Casewell}, {Decin}, {Dos Santos}, {Garc{\'\i}a Mu{\~n}oz}, {Harrington}, {Heng}, {Hu}, {Mancini}, {Molaverdikhani}, {Morello}, {Nikolov}, {Nixon}, {Redfield}, {Stevenson}, {Wakeford}, {Alam}, {Benneke}, {Blecic}, {Crouzet}, {Daylan}, {Inglis}, {Kreidberg}, {Petit dit de la Roche}, \& {Turner}}]{Esparza-Borges2023}
{Esparza-Borges}, E., {L{\'o}pez-Morales}, M., {Adams Redai}, J.~I., {et~al.} 2023, \apjl, 955, L19, \dodoi{10.3847/2041-8213/acf27b}

\bibitem[{{France} {et~al.}(2016){France}, {Loyd}, {Youngblood}, {Brown}, {Schneider}, {Hawley}, {Froning}, {Linsky}, {Roberge}, {Buccino}, {Davenport}, {Fontenla}, {Kaltenegger}, {Kowalski}, {Mauas}, {Miguel}, {Redfield}, {Rugheimer}, {Tian}, {Vieytes}, {Walkowicz}, \& {Weisenburger}}]{France2016}
{France}, K., {Loyd}, R.~O.~P., {Youngblood}, A., {et~al.} 2016, \apj, 820, 89, \dodoi{10.3847/0004-637X/820/2/89}

\bibitem[{Gueymard(2018)}]{Gueymard2018}
Gueymard, C.~A. 2018, Solar Energy, 169, 434 , \dodoi{https://doi.org/10.1016/j.solener.2018.04.067}

\bibitem[{{Hammond} {et~al.}(2020){Hammond}, {Tsai}, \& {Pierrehumbert}}]{Hammond2020}
{Hammond}, M., {Tsai}, S.-M., \& {Pierrehumbert}, R.~T. 2020, \apj, 901, 78, \dodoi{10.3847/1538-4357/abb08b}

\bibitem[{{Hauglustaine} {et~al.}(1994){Hauglustaine}, {Granier}, {Brasseur}, \& {M{\'e}Gie}}]{Hauglustaine1994}
{Hauglustaine}, D.~A., {Granier}, C., {Brasseur}, G.~P., \& {M{\'e}Gie}, G. 1994, \jgr, 99, 1173, \dodoi{10.1029/93JD02987}

\bibitem[{{House} {et~al.}(2003){House}, {Runnegar}, \& {Fitz-Gibbon}}]{House2003}
{House}, C.~H., {Runnegar}, B., \& {Fitz-Gibbon}, S.~T. 2003, Geobiology, 1, 15, \dodoi{10.1046/j.1472-4669.2003.00004.x}

\bibitem[{{Howard} {et~al.}(2012){Howard}, {Marcy}, {Bryson}, {Jenkins}, {Rowe}, {Batalha}, {Borucki}, {Koch}, {Dunham}, {Gautier}, {Van Cleve}, {Cochran}, {Latham}, {Lissauer}, {Torres}, {Brown}, {Gilliland}, {Buchhave}, {Caldwell}, {Christensen-Dalsgaard}, {Ciardi}, {Fressin}, {Haas}, {Howell}, {Kjeldsen}, {Seager}, {Rogers}, {Sasselov}, {Steffen}, {Basri}, {Charbonneau}, {Christiansen}, {Clarke}, {Dupree}, {Fabrycky}, {Fischer}, {Ford}, {Fortney}, {Tarter}, {Girouard}, {Holman}, {Johnson}, {Klaus}, {Machalek}, {Moorhead}, {Morehead}, {Ragozzine}, {Tenenbaum}, {Twicken}, {Quinn}, {Isaacson}, {Shporer}, {Lucas}, {Walkowicz}, {Welsh}, {Boss}, {Devore}, {Gould}, {Smith}, {Morris}, {Prsa}, {Morton}, {Still}, {Thompson}, {Mullally}, {Endl}, \& {MacQueen}}]{Howard2012}
{Howard}, A.~W., {Marcy}, G.~W., {Bryson}, S.~T., {et~al.} 2012, \apjs, 201, 15, \dodoi{10.1088/0067-0049/201/2/15}

\bibitem[{Hu {et~al.}(2021)Hu, Damiano, Scheucher, Kite, Seager, \& Rauer}]{Hu2021}
Hu, R., Damiano, M., Scheucher, M., {et~al.} 2021, 921, L8, \dodoi{10.3847/2041-8213/ac1f92}

\bibitem[{Hu {et~al.}(2012)Hu, Seager, \& Bains}]{Hu2012}
Hu, R., Seager, S., \& Bains, W. 2012, Astrophys. J., 761, \dodoi{10.1088/0004-637X/761/2/166}

\bibitem[{{Innes} \& {Pierrehumbert}(2022)}]{Innes2022}
{Innes}, H., \& {Pierrehumbert}, R.~T. 2022, \apj, 927, 38, \dodoi{10.3847/1538-4357/ac4887}

\bibitem[{{Innes} {et~al.}(2023){Innes}, {Tsai}, \& {Pierrehumbert}}]{Innes2023}
{Innes}, H., {Tsai}, S.-M., \& {Pierrehumbert}, R.~T. 2023, \apj, 953, 168, \dodoi{10.3847/1538-4357/ace346}

\bibitem[{{Johnson} {et~al.}(2004){Johnson}, {Sams}, \& {Sharpe}}]{PNNL2004}
{Johnson}, T.~J., {Sams}, R.~L., \& {Sharpe}, S.~W. 2004, in Society of Photo-Optical Instrumentation Engineers (SPIE) Conference Series, Vol. 5269, Chemical and Biological Point Sensors for Homeland Defense, ed. I.~{Sedlacek}, Arthur~J., R.~{Colton}, \& T.~{Vo-Dinh}, 159--167, \dodoi{10.1117/12.515604}

\bibitem[{Judeikis \& {G. Wren}(1977)}]{Judeikis1977}
Judeikis, H.~S., \& {G. Wren}, A. 1977, Atmospheric Environment (1967), 11, 1221, \dodoi{https://doi.org/10.1016/0004-6981(77)90099-3}

\bibitem[{{Kettle} {et~al.}(2001){Kettle}, {Rhee}, {von Hobe}, {Poulton}, {Aiken}, \& {Andreae}}]{Kettle2001}
{Kettle}, A.~J., {Rhee}, T.~S., {von Hobe}, M., {et~al.} 2001, \jgr, 106, 12,193, \dodoi{10.1029/2000JD900630}

\bibitem[{{Kite} \& {Ford}(2018)}]{Kite2018}
{Kite}, E.~S., \& {Ford}, E.~B. 2018, \apj, 864, 75, \dodoi{10.3847/1538-4357/aad6e0}

\bibitem[{Klippenstein(2023)}]{Klippenstein2023}
Klippenstein, S.~J. 2023, {Private Communication}

\bibitem[{{Leconte} {et~al.}(2024){Leconte}, {Spiga}, {Cl{\'e}ment}, {Guerlet}, {Selsis}, {Milcareck}, {Cavali{\'e}}, {Moreno}, {Lellouch}, {Carri{\'o}n-Gonz{\'a}lez}, {Charnay}, \& {Lef{\`e}vre}}]{Leconte2024}
{Leconte}, J., {Spiga}, A., {Cl{\'e}ment}, N., {et~al.} 2024, arXiv e-prints, arXiv:2401.06608, \dodoi{10.48550/arXiv.2401.06608}

\bibitem[{Lee {et~al.}(2021)Lee, Parmentier, Hammond, Grimm, Kitzmann, Tan, Tsai, \& Pierrehumbert}]{Lee2021}
Lee, E. K.~H., Parmentier, V., Hammond, M., {et~al.} 2021, MNRAS, 506, 2695.
\newblock \doarXiv{2106.11664}

\bibitem[{Lee {et~al.}(2020)Lee, Casewell, Chubb, Hammond, Tan, Tsai, \& Pierrehumbert}]{Lee2020}
Lee, G.~K., Casewell, S.~L., Chubb, K.~L., {et~al.} 2020, Monthly Notices of the Royal Astronomical Society, 496, 4674, \dodoi{10.1093/mnras/staa1882}

\bibitem[{{Leung} {et~al.}(2022){Leung}, {Schwieterman}, {Parenteau}, \& {Fauchez}}]{Leung2022}
{Leung}, M., {Schwieterman}, E.~W., {Parenteau}, M.~N., \& {Fauchez}, T.~J. 2022, \apj, 938, 6, \dodoi{10.3847/1538-4357/ac8799}

\bibitem[{Li {et~al.}(2023)Li, Cao, Wang, Carrión, Zhu, Miao, Wang, Chen, Todd, \& Zhang}]{Li2023}
Li, C.-Y., Cao, H.-Y., Wang, Q., {et~al.} 2023, The ISME Journal, 17, 1184, \dodoi{10.1038/s41396-023-01430-z}

\bibitem[{{Lovelock}(1989)}]{Lovelock1989}
{Lovelock}, J.~E. 1989, Reviews of Geophysics, 27, 215, \dodoi{10.1029/RG027i002p00215}

\bibitem[{{Loyd} {et~al.}(2016){Loyd}, {France}, {Youngblood}, {Schneider}, {Brown}, {Hu}, {Linsky}, {Froning}, {Redfield}, {Rugheimer}, \& {Tian}}]{Loyd2016}
{Loyd}, R.~O.~P., {France}, K., {Youngblood}, A., {et~al.} 2016, \apj, 824, 102, \dodoi{10.3847/0004-637X/824/2/102}

\bibitem[{{Luque} \& {Pall{\'e}}(2022)}]{Luque2022}
{Luque}, R., \& {Pall{\'e}}, E. 2022, Science, 377, 1211, \dodoi{10.1126/science.abl7164}

\bibitem[{{Madhusudhan} {et~al.}(2020){Madhusudhan}, {Nixon}, {Welbanks}, {Piette}, \& {Booth}}]{Madhusudhan2020}
{Madhusudhan}, N., {Nixon}, M.~C., {Welbanks}, L., {Piette}, A. A.~A., \& {Booth}, R.~A. 2020, \apjl, 891, L7, \dodoi{10.3847/2041-8213/ab7229}

\bibitem[{{Madhusudhan} {et~al.}(2021){Madhusudhan}, {Piette}, \& {Constantinou}}]{Madhusudhan2021}
{Madhusudhan}, N., {Piette}, A. A.~A., \& {Constantinou}, S. 2021, \apj, 918, 1, \dodoi{10.3847/1538-4357/abfd9c}

\bibitem[{{Madhusudhan} {et~al.}(2023){Madhusudhan}, {Sarkar}, {Constantinou}, {Holmberg}, {Piette}, \& {Moses}}]{Madhusudhan2023}
{Madhusudhan}, N., {Sarkar}, S., {Constantinou}, S., {et~al.} 2023, \apjl, 956, L13, \dodoi{10.3847/2041-8213/acf577}

\bibitem[{{Malik} {et~al.}(2019b){Malik}, {Kempton}, {Koll}, {Mansfield}, {Bean}, \& {Kite}}]{Malik2019b}
{Malik}, M., {Kempton}, E. M.~R., {Koll}, D. D.~B., {et~al.} 2019b, \apj, 886, 142, \dodoi{10.3847/1538-4357/ab4a05}

\bibitem[{Malik {et~al.}(2019a)Malik, Kitzmann, Mendon{\c{c}}a, Grimm, Marleau, Linder, Tsai, \& Heng}]{Malik2019a}
Malik, M., Kitzmann, D., Mendon{\c{c}}a, J.~M., {et~al.} 2019a, The Astronomical Journal, 157, 170, \dodoi{10.3847/1538-3881/ab1084}

\bibitem[{Mateos {et~al.}(2023)Mateos, Chappell, Klos, Le, Boden, Stüeken, \& Anderson}]{Mateos2023}
Mateos, K., Chappell, G., Klos, A., {et~al.} 2023, Science Advances, 9, eade4847, \dodoi{10.1126/sciadv.ade4847}

\bibitem[{{Meadows} {et~al.}(2023){Meadows}, {Lincowski}, \& {Lustig-Yaeger}}]{Meadows2023}
{Meadows}, V., {Lincowski}, A., \& {Lustig-Yaeger}, J. 2023, in American Astronomical Society Meeting Abstracts, Vol.~55, American Astronomical Society Meeting Abstracts, 125.04

\bibitem[{{Nixon} \& {Madhusudhan}(2021)}]{Nixon2021}
{Nixon}, M.~C., \& {Madhusudhan}, N. 2021, \mnras, 505, 3414, \dodoi{10.1093/mnras/stab1500}

\bibitem[{{Peacock} {et~al.}(2020){Peacock}, {Barman}, {Shkolnik}, {Loyd}, {Schneider}, {Pagano}, \& {Meadows}}]{Peacock2020}
{Peacock}, S., {Barman}, T., {Shkolnik}, E.~L., {et~al.} 2020, \apj, 895, 5, \dodoi{10.3847/1538-4357/ab893a}

\bibitem[{{Pierrehumbert}(2023)}]{Pierrehumbert2023}
{Pierrehumbert}, R.~T. 2023, \apj, 944, 20, \dodoi{10.3847/1538-4357/acafdf}

\bibitem[{{Piette} \& {Madhusudhan}(2020)}]{Piette2020}
{Piette}, A. A.~A., \& {Madhusudhan}, N. 2020, \apj, 904, 154, \dodoi{10.3847/1538-4357/abbfb1}

\bibitem[{Pilcher(2003)}]{Pilcher2003}
Pilcher, C.~B. 2003, Astrobiology, 3, 471, \dodoi{10.1089/153110703322610582}

\bibitem[{{Rogers} {et~al.}(2023){Rogers}, {Schlichting}, \& {Owen}}]{Rogers2023}
{Rogers}, J.~G., {Schlichting}, H.~E., \& {Owen}, J.~E. 2023, \apjl, 947, L19, \dodoi{10.3847/2041-8213/acc86f}

\bibitem[{Schwieterman {et~al.}(2018)Schwieterman, Kiang, Parenteau, Harman, DasSarma, Fisher, Arney, Hartnett, Reinhard, Olson, Meadows, Cockell, Walker, Grenfell, Hegde, Rugheimer, Hu, \& Lyons}]{Schwieterman2018}
Schwieterman, E.~W., Kiang, N.~Y., Parenteau, M.~N., {et~al.} 2018, Astrobiology, 18, 663, \dodoi{10.1089/ast.2017.1729}

\bibitem[{{Seager} {et~al.}(2013){Seager}, {Bains}, \& {Hu}}]{Seager2013}
{Seager}, S., {Bains}, W., \& {Hu}, R. 2013, \apj, 777, 95, \dodoi{10.1088/0004-637X/777/2/95}

\bibitem[{Seinfeld \& Pandis(2016)}]{Seinfeld2016}
Seinfeld, J.~H., \& Pandis, S.~N. 2016, Atmospheric chemistry and physics: from air pollution to climate change (John Wiley \&amp; Sons, Inc.)

\bibitem[{Smith(1998)}]{Smith1998}
Smith, M.~D. 1998, Icarus, 132, 176 , \dodoi{https://doi.org/10.1006/icar.1997.5886}

\bibitem[{{Thompson} {et~al.}(2018){Thompson}, {Coughlin}, {Hoffman}, {Mullally}, {Christiansen}, {Burke}, {Bryson}, {Batalha}, {Haas}, {Catanzarite}, {Rowe}, {Barentsen}, {Caldwell}, {Clarke}, {Jenkins}, {Li}, {Latham}, {Lissauer}, {Mathur}, {Morris}, {Seader}, {Smith}, {Klaus}, {Twicken}, {Van Cleve}, {Wohler}, {Akeson}, {Ciardi}, {Cochran}, {Henze}, {Howell}, {Huber}, {Pr{\v{s}}a}, {Ram{\'\i}rez}, {Morton}, {Barclay}, {Campbell}, {Chaplin}, {Charbonneau}, {Christensen-Dalsgaard}, {Dotson}, {Doyle}, {Dunham}, {Dupree}, {Ford}, {Geary}, {Girouard}, {Isaacson}, {Kjeldsen}, {Quintana}, {Ragozzine}, {Shabram}, {Shporer}, {Silva Aguirre}, {Steffen}, {Still}, {Tenenbaum}, {Welsh}, {Wolfgang}, {Zamudio}, {Koch}, \& {Borucki}}]{Thompson2018}
{Thompson}, S.~E., {Coughlin}, J.~L., {Hoffman}, K., {et~al.} 2018, \apjs, 235, 38, \dodoi{10.3847/1538-4365/aab4f9}

\bibitem[{{Tsai} {et~al.}(2014){Tsai}, {Dobbs-Dixon}, \& {Gu}}]{Tsai2014}
{Tsai}, S.-M., {Dobbs-Dixon}, I., \& {Gu}, P.-G. 2014, \apj, 793, 141, \dodoi{10.1088/0004-637X/793/2/141}

\bibitem[{Tsai {et~al.}(2021b)Tsai, Innes, Lichtenberg, Taylor, Malik, Chubb, \& Pierrehumbert}]{Tsai2021b}
Tsai, S.-M., Innes, H., Lichtenberg, T., {et~al.} 2021b, The Astrophysical Journal Letters, 922, L27, \dodoi{10.3847/2041-8213/ac399a}

\bibitem[{Tsai {et~al.}(2018)Tsai, Kitzmann, Lyons, Mendon{\c{c}}a, Grimm, \& Heng}]{Tsai2018}
Tsai, S.-M., Kitzmann, D., Lyons, J.~R., {et~al.} 2018, ApJ, 862, 31, \dodoi{10.3847/1538-4357/aac834}

\bibitem[{Tsai {et~al.}(2017)Tsai, Lyons, Grosheintz, Rimmer, Kitzmann, \& Heng}]{Tsai2017}
Tsai, S.-M., Lyons, J.~R., Grosheintz, L., {et~al.} 2017, Astrophys. J. Suppl. Ser., 228, 1, \dodoi{10.3847/1538-4365/228/2/20}

\bibitem[{Tsai {et~al.}(2021)Tsai, Malik, Kitzmann, Lyons, Fateev, Lee, \& Heng}]{Tsai2021}
Tsai, S.-M., Malik, M., Kitzmann, D., {et~al.} 2021, The Astrophysical Journal, 923, 264, \dodoi{10.3847/1538-4357/ac29bc}

\bibitem[{{Tsai} {et~al.}(2023c){Tsai}, {Moses}, {Powell}, \& {Lee}}]{Tsai2023c}
{Tsai}, S.-M., {Moses}, J.~I., {Powell}, D., \& {Lee}, E. K.~H. 2023c, \apjl, 959, L30, \dodoi{10.3847/2041-8213/ad1405}

\bibitem[{{Tsai} {et~al.}(2024){Tsai}, {Parmentier}, {Mendon{\c{c}}a}, {Tan}, {Deitrick}, {Hammond}, {Savel}, {Zhang}, {Pierrehumbert}, \& {Schwieterman}}]{Tsai2024}
{Tsai}, S.-M., {Parmentier}, V., {Mendon{\c{c}}a}, J.~M., {et~al.} 2024, arXiv e-prints, arXiv:2310.17751, \dodoi{10.48550/arXiv.2310.17751}

\bibitem[{Tsang(1987)}]{Tsang1987}
Tsang, W. 1987, Journal of Physical and Chemical Reference Data, 16, 471, \dodoi{10.1063/1.555802}

\bibitem[{Tsang \& Hampson(1986)}]{Tsang1986}
Tsang, W., \& Hampson, R.~F. 1986, Journal of Physical and Chemical Reference Data, 15, 1087, \dodoi{10.1063/1.555759}

\bibitem[{{Tsiaras} {et~al.}(2019){Tsiaras}, {Waldmann}, {Tinetti}, {Tennyson}, \& {Yurchenko}}]{Tsiaras2019}
{Tsiaras}, A., {Waldmann}, I.~P., {Tinetti}, G., {Tennyson}, J., \& {Yurchenko}, S.~N. 2019, Nature Astronomy, 3, 1086, \dodoi{10.1038/s41550-019-0878-9}

\bibitem[{Tsuboi {et~al.}(1981)Tsuboi, Katoh, Kikuchi, \& Hashimoto}]{Tsuboi1981}
Tsuboi, T., Katoh, M., Kikuchi, S., \& Hashimoto, K. 1981, Japanese Journal of Applied Physics, 20, 985, \dodoi{10.1143/JJAP.20.985}

\bibitem[{van~der Walt {et~al.}(2011)van~der Walt, Colbert, \& Varoquaux}]{Walt2011}
van~der Walt, S., Colbert, S.~C., \& Varoquaux, G. 2011, Computing in Science Engineering, 13, 22, \dodoi{10.1109/MCSE.2011.37}

\bibitem[{{Venturini} {et~al.}(2020){Venturini}, {Guilera}, {Haldemann}, {Ronco}, \& {Mordasini}}]{Venturini2020}
{Venturini}, J., {Guilera}, O.~M., {Haldemann}, J., {Ronco}, M.~P., \& {Mordasini}, C. 2020, \aap, 643, L1, \dodoi{10.1051/0004-6361/202039141}

\bibitem[{{Wogan} {et~al.}(2024){Wogan}, {Batalha}, {Zahnle}, {Krissansen-Totton}, {Tsai}, \& {Hu}}]{Wogan2024}
{Wogan}, N.~F., {Batalha}, N.~E., {Zahnle}, K.~J., {et~al.} 2024, \apjl, 963, L7, \dodoi{10.3847/2041-8213/ad2616}

\bibitem[{{Xu} {et~al.}(2015){Xu}, {Raghunath}, \& {Lin}}]{Xu2015}
{Xu}, Z.~F., {Raghunath}, P., \& {Lin}, M.~C. 2015, Journal of Physical Chemistry A, 119, 7404, \dodoi{10.1021/acs.jpca.5b00553}

\bibitem[{{Youngblood} {et~al.}(2016){Youngblood}, {France}, {Loyd}, {Linsky}, {Redfield}, {Schneider}, {Wood}, {Brown}, {Froning}, {Miguel}, {Rugheimer}, \& {Walkowicz}}]{Youngblood2016}
{Youngblood}, A., {France}, K., {Loyd}, R.~O.~P., {et~al.} 2016, \apj, 824, 101, \dodoi{10.3847/0004-637X/824/2/101}

\bibitem[{{Yu} {et~al.}(2021){Yu}, {Moses}, {Fortney}, \& {Zhang}}]{Yu2021}
{Yu}, X., {Moses}, J.~I., {Fortney}, J.~J., \& {Zhang}, X. 2021, \apj, 914, 38, \dodoi{10.3847/1538-4357/abfdc7}

\bibitem[{{Zeng} {et~al.}(2019){Zeng}, {Jacobsen}, {Sasselov}, {Petaev}, {Vanderburg}, {Lopez-Morales}, {Perez-Mercader}, {Mattsson}, {Li}, {Heising}, {Bonomo}, {Damasso}, {Berger}, {Cao}, {Levi}, \& {Wordsworth}}]{Zeng2019}
{Zeng}, L., {Jacobsen}, S.~B., {Sasselov}, D.~D., {et~al.} 2019, Proceedings of the National Academy of Science, 116, 9723, \dodoi{10.1073/pnas.1812905116}

\bibitem[{{Zhang} {et~al.}(2019){Zhang}, {Chachan}, {Kempton}, \& {Knutson}}]{Zhang2019}
{Zhang}, M., {Chachan}, Y., {Kempton}, E. M.~R., \& {Knutson}, H.~A. 2019, \pasp, 131, 034501, \dodoi{10.1088/1538-3873/aaf5ad}

\bibitem[{{Zhang} {et~al.}(2020){Zhang}, {Chachan}, {Kempton}, {Knutson}, \& {Chang}}]{Zhang2020}
{Zhang}, M., {Chachan}, Y., {Kempton}, E. M.~R., {Knutson}, H.~A., \& {Chang}, W.~H. 2020, \apj, 899, 27, \dodoi{10.3847/1538-4357/aba1e6}

\bibitem[{Zhang {et~al.}(2020)Zhang, Park, Yan, Park, Wu, Jang, Gao, Tan, Wang, \& Chen}]{earthDMS}
Zhang, M., Park, K.-T., Yan, J., {et~al.} 2020, Progress in Oceanography, 186, 102392, \dodoi{https://doi.org/10.1016/j.pocean.2020.102392}

\end{thebibliography}


\end{document}